\documentclass[a4paper,11pt]{article}
\pdfoutput=1
\usepackage{jcappub}

\usepackage{tikz,xcolor,hyperref}
\usepackage{amsmath, amssymb, amsthm, graphicx, epsfig, fancyhdr,epsfig, slashed}
\usepackage{mathtools}
\usepackage[normalem]{ulem}
\usepackage{tikzsymbols}
\usepackage{natbib}
\usepackage{float}
\usepackage{amsmath}
\usepackage{amssymb}
\usepackage{amsfonts}

\newcommand{\gs}{g_\star}
\newcommand{\gss}{g_{\star s}}
\newcommand{\Trh}{T_\text{rh}}

\newcommand{\Tmax}{T_\text{max}}

\newcommand{\mdm}{m_{\rm DM}}

\begin{document}
\title{Light PIDM in Warped Extra Dimensions }
\author[a]{Mathew Thomas Arun,}
\emailAdd{mathewthomas@iisertvm.ac.in }
\author[a]{Prachiti P. Athalye,}
\emailAdd{prachiti23@iisertvm.ac.in}
\author[b]{and Basabendu Barman}
\emailAdd{basabendu.b@srmap.edu.in}
\affiliation[a]{\,\,School of Physics, Indian Institute of Science Education and Research, Thiruvananthapuram-695551, Kerala, India}
\affiliation[b]{\,\,Department of Physics, School of Engineering and Sciences, SRM University-AP, Amaravati 522240, India}
\abstract{
Traditional Planckian Interacting Dark Matter (PIDM), which interacts exclusively through gravity, typically requires heavy DM candidates (with mass $10^3-10^{15}$ GeV) and very high reheating temperature ($T_{\rm rh}\gtrsim 10^{15}$ GeV). In this article, we explore a novel realization of PIDM in warped five-dimensions, consisting of an {\it Ultra Violet}--{\it Dark}--{\it Infra Red} (UV-DB-IR) brane setup, where the DM can be a {\it Dark} brane composite light state with mass 1 MeV\---1 TeV. The DM sector is assumed to interact solely via gravity in five-dimensions. After orbifolding and performing a Kaluza-Klein (KK) decomposition, the DM is assumed to be localized onto the DB, which is positioned in the extra-dimension such that the DM interacts with both the massless graviton and its massive KK excitations, with suppressed couplings to remain consistent with the ethos of the PIDM framework. The light (heavy) Standard Model matter is assumed to be localized near UV (IR) branes for the geometric Froggatt-Neilsen mechanism, while their KK modes are localized close to the IR brane. We show that this construction allows for a viable and efficient freeze-in production mechanism for light composite PIDM, consistent with TeV-scale reheating temperature.
}
\maketitle
\section{Introduction}
\label{sec:intro}
Baryonic matter constitutes only about 4\% of the total matter-energy content of the Universe~\cite{Planck:2018vyg}, while approximately 26\% is attributed to dark matter (DM)~\cite{Jungman:1995df, Bertone:2016nfn, deSwart:2017heh}, which stands as a longstanding and fundamental mystery in both particle physics and cosmology. Stringent observational constraints, particularly from direct detection experiments, have significantly narrowed the parameter space of conventional weakly interacting massive particles (WIMPs), the leading class of DM candidates (see, e.g., Refs.~\cite{Roszkowski:2017nbc, Arcadi:2017kky, Arcadi:2024ukq}). This has motivated the investigation of alternative DM production mechanisms. A well-studied alternative is the feebly interacting massive particle (FIMP) framework, in which DM is produced via the decay or annihilation of particles in the visible sector in the early Universe. As the temperature of the Standard Model (SM) plasma falls below the relevant interaction scale, DM production becomes Boltzmann suppressed, resulting in a constant comoving number density---a process known as freeze-in~\cite{McDonald:2001vt, Hall:2009bx, Bernal:2017kxu}. The FIMP scenario necessitates extremely suppressed interactions between the dark and visible sectors to ensure non-thermal production. Such interactions can arise either from small infrared couplings or from higher-dimensional, non-renormalizable operators suppressed by a large new physics (NP) scale, as in ultraviolet freeze-in (UVFI)~\cite{Hall:2009bx, Elahi:2014fsa, Barman:2020plp}. Even feebler interactions arise when gravity constitutes the sole effective coupling at the UV scale. In this context, the most minimal and unavoidable interaction between DM and the SM is mediated by graviton exchange~\cite{Ema:2015dka,Garny:2015sjg,Tang:2016vch,Ema:2016hlw,Garny:2017kha,Tang:2017hvq,Bernal:2018qlk,Ema:2018ucl,Ema:2019yrd,Redi:2020ffc,Chianese:2020yjo,Chianese:2020khl,Mambrini:2021zpp,Barman:2021ugy,Haque:2021mab,Clery:2021bwz,Clery:2022wib,Ahmed:2022tfm} (see~\cite{Kolb:2023ydq} and the references therein for a recent review), potentially yielding the observed DM abundance through scatterings among particles within the thermal plasma. Such models have recently gathered traction because of two reasons: (a) the Large Hadron Collider (LHC) has not found any evidence of New Physics yet, indicating the existence of only four fundamental forces, and (b) DM interacts purely via gravity. On the other hand, the `naturalness' becomes an issue if we assume that there lies a desert between the electroweak and the Planck scale. To motivate a high (Planck) scale suppressed effective interaction with gravity alone for the DM, it might be worthwhile to assume a `co-genesis' for DM and graviton.

However, it must be emphasized that such a model, namely Planckian Interacting Dark Matter (PIDM), usually requires a heavy DM to satisfy the relic density constraint (See, for example, Refs.~\cite{Garny:2015sjg,Garny:2017kha}) along with a GUT scale reheating. This is contrary to our co-genesis motivation, which suggests that the mass of DM at best must be soft if it shares a common origin with the graviton, which is massless and potentially a Goldstone boson~\cite{Ohanian:1969xhl,Phillips:1966zzc, Chkareuli:2001xe, Berezhiani:2007zf,Berezhiani:2008ue,Carroll:2009mr,Tomboulis:2011qh}. On the other hand, this kind of DM would not satisfy the relic density constraint, as they are too light (with a mass $\sim\mathcal{O}(10^{-3}-10^2)\,\text{GeV}$),  and would require a reheating temperature $\Trh\sim M_P$. Here, we address this gap in the formalism while keeping to the ethics of Planckian suppressed coupling for DM. This is distinctly different from the scenarios discussed, like in~\cite{Lee:2013bua,Lee:2024wes,Bernal:2020fvw,Bernal:2020yqg}, that assume a common origin and universal couplings to the dark and electroweak sectors. Though in~\cite{Lee:2013bua,Lee:2024wes} the authors discuss composite/partial composite DM, the {\it Dark} brane is located together with the IR brane ensuring that the DM, via $\sim\mathcal{O}(\text{TeV}^{-1})$ coupling with KK graviton, becomes thermal. Whereas, we are interested in the scenario in which the DM always has $\sim\mathcal{O}(M_P^{-1})$ interaction with all KK graviton states.
\begin{figure}[htb!]
    \centering    
    \includegraphics[scale=0.32]{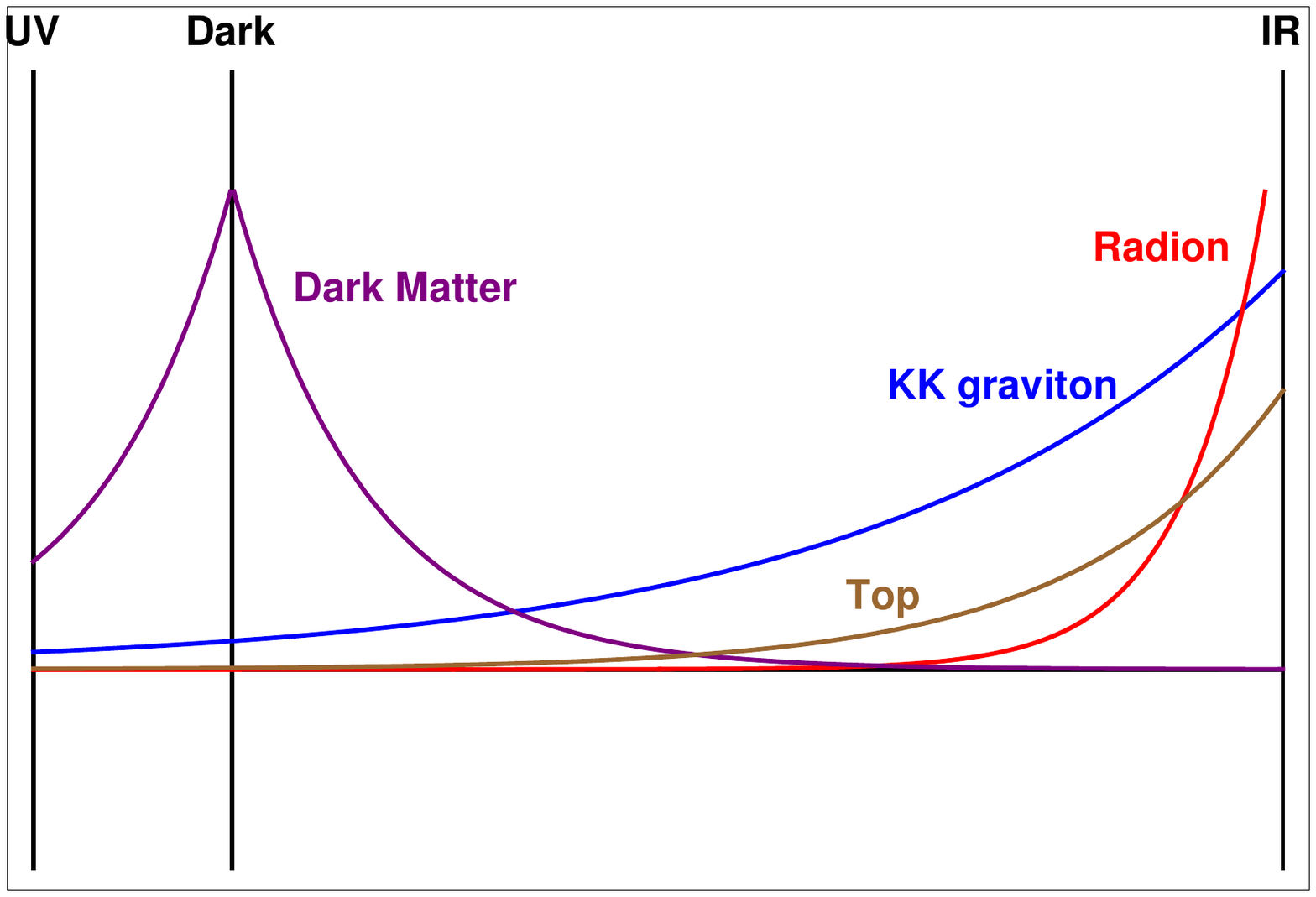}
    \caption{Schematic diagram of the present model in the warped geometry. Here ``UV'', ``Dark'' and ``IR'' indicate the locations of the UV-brane, dark-brane and IR-brane, respectively.}
    \label{fig:schematic}
\end{figure}

In this work, instead of addressing the UV symmetry breaking and their features, we will focus on the DM phenomenology by considering a low-energy version of the model in which the light composite DM field is localized on the dark brane, as shown in Fig.~\ref{fig:schematic}, in a five-dimensional warped Randall-Sundrum model. Since the energy scales we explore are $\sim \mathcal{O}(\text{TeV})$, much below the Planck scale, we are safe in considering the composite DM as scalar, vector or fermion field, just like below the chiral symmetry breaking scale the pion, $\rho$-meson and proton are understood by scalar, vector and fermion. The reason we consider a five-dimensional warped scenario is that this is the only `natural' mechanism to generate a substantial relic for our light composite DM, taking into account the effects of Kaluza-Klein (KK) towers of the graviton, which enhance non-thermal production. Unlike flat five-dimensional $S^1/Z_2$ models~\cite{Garny:2017kha}, KK partners of gravitons have $\mathcal{O}(1)$ coupling with IR localized fields. The dark brane is assumed to be near the UV brane, to explain the graviton-DM origin, making sure that the effective coupling of DM remains highly suppressed with all the KK graviton modes. While this placing of dark brane motivates a common origin, as a low-energy effective theory, we consider two distinct types of DM localizations. The first is, like Higgs, a composite DM which is entirely confined to the dark brane, namely dark brane composite DM, while the second where a partial composite DM where the bulk wave profile, though is localized to the dark brane does also permeate into the bulk, namely the {dark brane {\it partially} composite DM. These scenarios lead to distinct phenomenology as,
\begin{itemize}
\item[(i)] {\bf Dark brane composite DM}: Like Higgs on the IR brane, if we consider the DM to be localized entirely on the dark brane, they will not have any heavier KK modes. The positioning of the dark brane in the bulk would determine the coupling of DM with the KK gravitons. For them to be PIDM, we assume the dark brane positioning such that the couplings are Planck suppressed. The non-thermal production of these DM candidates would be dominated by $F+F \to {\rm DM} +{\rm DM}$ via KK graviton exchange diagrams, where $F$ can be a right handed top or any KK partner of SM fields in the thermal bath. 
\item[(ii)] {\bf Dark brane partially composite DM}: On the other hand, we may also consider the DM to be a bulk field localized on the dark brane. In this scenario, the heavier KK modes will become important and these heavier modes of the DM, localized towards the IR, can be in the thermal bath, and produced via freeze-out. Once produced, they produce the zero-modes via 2-to-2 graviton-mediated annihilations. These stable non-decaying zero modes constitute the total DM abundance.
\end{itemize}
To satisfy the fermion mass hierarchy, the geometric Froggatt-Nielsen mechanism~\cite{Froggatt:1978nt} requires localization of light fermions onto the UV brane, while the right handed top to the IR one. The warped scenario is distinct that, while the SM gauge and graviton fields are non-localized in the bulk, being the zero mode, their KK partners exhibit partial compositeness. Since that the DM considered has Planckian suppressed interactions with all KK modes of graviton, interesting non-thermal production channels arise from the annihilation of right handed top quark and their KK towers and the KK towers of gauge bosons via KK gravitons.

The article is organized as follows. In Sec.~\ref{sec:extdim}, we introduce the Randall-Sundrum model with bulk fermions localized to UV and IR branes to satisfy their mass hierarchy. Hence, the non-thermal production of DM will contain the annihilation and decays of heavy quarks, Higgs and other KK modes. We review the field localizations in the geometry and their coupling strengths with gravitons for completeness. Towards the end, we also discuss the current direct and indirect bounds on warped extra-dimension set-up. In Sec.~\ref{sec:dm}, we compute the freeze-in production for both {\it Dark} brane composite and dark brane partially composite DM models. These scenarios are distinct, while the former does not contain any decay processes from thermal KK partners of DM, the latter does. Finally, we conclude in Sec.~\ref{sec:concl} by summarizing our findings. 
\section{Extra dimensional set-up}
\label{sec:extdim}
Here we will briefly recap the Randall-Sundrum model~\cite{Randall:1999ee,Randall:1999vf}, to make the article self contained. Consider a $S^1/Z_2$ orbifolded 5-dimensions with 4 space and one time dimensions, warped along the $5^{\rm th}$ dimension. The geometry is defined by the line element, 
\begin{align}
ds^2 = g_{MN}\,dx^M\,dx^N = e^{-kr_c |y|}\,\eta_{\mu \nu }\,dx^{\mu}\,dx^{\nu} - r_c^2\,dy^2\,,    
\end{align}
where $\eta_{\mu \nu} = {\rm diag}(-1,+1,+1,+1)$ and $e^{-kr_c |y|}$ is the warping with $0 \leq y \leq \pi$. The extra dimension is a line-segment $\phi = yr_c \in (0,\pi r_c)$. The two orbifold fixed points, $y=0$ and $y=\pi$, are protected by two 4-dimensional membranes, called UV and the IR brane. The warp factor $k\sim 11$ is set to solve the Higgs mass hierarchy problem, preventing the Higgs from getting large uncontrolled mass corrections. This forms a `natural' setting for the study of `Planckian' New Physics, with a fundamental Planck scale $M_5 \sim \mathcal{O}(10^{18}\,\text{GeV})$. We assume the bulk to be populated by gauge and fermion fields transforming under the adjoint and fundamental representations of the SM gauge group $SU(3)_c \times SU(2)_W \times U(1)_Y$. 
The Higgs field is assumed to be localized on the IR brane to stabilize the Higgs vacuum expectation value to $\langle H \rangle = M_5\,e^{-k r_c \pi}$. Upon compactification the bulk fields decompose into infinite states of Kaluza Klein (KK) towers, which are regulated using a cutoff. Here, we assume that the cutoff is given by $\Lambda/M_{\rm KK} = 4$, wherein we allow the first four KK modes of the fields. 

The fermion content in the bulk contains three copies of quark doublets and singlets each and similarly for the lepton sector. Unlike four-dimensions, here the Clifford algebra is not reducible. This leads to doubling of degrees of freedom, at each Kaluza Klein level, under compactification, while orbifolding boundary conditions make sure that the unwanted chiral modes are projected out and only the SM fermion chiral modes are present at the lightest KK level. These zero modes are localized onto UV or IR brane depending on their mass term in the bulk action. The higher KK towers, on the other hand, all localize to the IR brane indicating partial compositeness among these states. Unlike the fermions, the  KK zero mode of graviton and gauge boson are not localized and have flat profile in the bulk. While, the non-zero KK modes all again localize on to the IR brane. Since the Higgs is localized at $y=\pi$, the electroweak symmetry breaking occurs on the IR brane, and the masses of fermion and electroweak gauge fields become directly proportional to the overlap of its wave profile in the extra-dimension with the Higgs profile. Thus, the Randall-Sundrum model generates a geometric Froggatt-Nielsen mechanism by explaining the large hierarchies in the fermion masses in SM with nearly anarchic bulk mass parameters. This means that except for the right handed top quark, the rest of the zero modes of fermions are localized towards the UV brane. 

In the present model we are interested in the coupling of graviton KK modes with Higgs and other bulk fields. By definition, all fields couple to KK zero graviton with Planck suppressed effective operator. On the other hand, since the KK partners of gravitons are localized to IR brane, all the heavy bulk fields will have substantial overlap leading to $\sim\mathcal{O}(\text{TeV}^{-1})$ suppressed effective coupling. Hence, in the current study, we consider the following visible sector fields which have the most overlap with KK modes of graviton:
\begin{itemize}
     \item [(i)] Zeroth mode of right handed top quark
     \item [(ii)] KK-1 modes of all fermion field
     \item [(iii)] KK-1 modes of all gauge field
     \item [(iv)] Higgs localized on the IR brane
\end{itemize}
For the dark sector localized on the dark brane, we explore two situations that can lead to interesting and distinct phenomenology as mentioned in the introduction, namely, the dark brane composite DM and the dark brane partially composite DM.
\subsection{Field profiles in the extra dimension}
Based on the previous discussion, we now  describe the wavefunction profiles for the scalar, fermion, gauge bosons, graviton and radion in the theory as follows,
\begin{itemize} 
\item {\underline {\it Scalar field}}:
Consider a bulk complex scalar field with action,
\begin{align}
    S_{\rm scalar} = \int d^5x \sqrt{-g}\Big( |\partial_M\Phi|^2 + m_{\Phi}^2 |\Phi|^2\Big) \ ,
\end{align}
where $m_\phi^2=a k^2$ is the bulk mass parameter defined in units of the warp factor $k$ and `$a$' is a dimensionless parameter. Varying this action, the resultant equation of motion is solved using the boundary conditions,
\begin{align}
    (\delta \Phi^\star\,\partial_5\Phi)|_{0,\pi r_c} = 0 \ .
\end{align}
Decomposing the five-dimensional field to its Fourier components,
\begin{align}
    \Phi(x^\mu,y) = \sum_{n=0}^{\infty} \Phi^{(n)}(x^\mu)f_\Phi^{n}(y) \ ,
\end{align}
where the extra-dimensional wave profile satisfies the orthonormality condition,
\begin{align}
    \int_0^{\pi r_c} dy e^{-2k r_c |y|} f_\Phi^{m} f_\Phi^{n} = \delta^{mn} \ .
\end{align}
With these, the solutions for the massless and the heavier KK modes become,
\begin{align}
& f_\Phi^{0}(y)=N_\Phi^0 e^{b k r_c y}\,, 
&
f_\Phi^{n}(y) = N_\Phi^n e^{2 k r_c y}\Big[ J_\alpha \Big(\frac{m_n}{k e^{- kr_c y}}\Big) + b_\Phi^{n} Y_\alpha\Big(\frac{m_n}{k e^{- kr_c y}}\Big) \Big]\,,
\end{align}
where $b=2\pm\sqrt{4+a}$ is related to the bulk mass parameter, while $N_\Phi^0$, $N_\Phi^n$ and $b_\Phi^{n}$, are arbitrary constants, fixed by boundary and orthonormality conditions.
The KK masses are determined by imposing the boundary conditions on the wave-profiles, and, are given by,
\begin{equation}
    m_n \approx \left( n + \frac{1}{2} \sqrt{4 + a} - \frac{3}{4} \right) \pi \ .
\end{equation}
\item  {\underline {\it Fermion field}}: The five-dimensional fermionic action for doublets ($Q$) and singlet ($q$) quarks with bulk mass terms is given as,
\begin{align}
& S_{\rm fermion} = \int d^5x \sqrt{-g}\,\left[\bar{Q}\,\left(\Gamma^M \partial_M + m_Q\right)\,Q + \bar{q}\,(\Gamma^M\partial_M + m_q)\,q + YH\bar{Q}\,q\,\delta(y-\pi)\right]\,.
\end{align}
The boundary conditions at the orbifold fixed points ($y=0, \ y=\pi$), to ensure the correct chiral modes are projected out at the lowest modes, are chosen to be,
\begin{equation}
    Q_L(++), \ Q_R(--), \ q_L(--), \ q_R(++) \ ,
\end{equation}
where $L, R$ stand for the left and right handed fields under the four-dimensional chiral projection operator. Decomposing the fields to their Fourier modes, upon compactification,
\begin{equation}
    q(x_\mu,y)_{L,R} = \sum_{n=0}^{\infty}\frac{1}{\sqrt{\pi r_c}}\, q_{L,R}^{(n)}(x_\mu)\,e^{2k r_c y}\,\hat{f}_{L,R}^{(n)}(y)\,,
\end{equation}
where $q_{L,R}^{(n)}$ stands for the four-dimensional chiral KK modes and $\hat{f}_{L,R}^{(n)}(y)$ are their wave-profiles in the bulk. These fields are set to satisfy the ortho-normality condition, 
\begin{align}
    \int_{-\pi}^\pi dy   e^{kr_c |y|} \hat{f}_L^{(m) *} \hat{f}_L^{(n)}=\int_{-\pi}^\pi d y e^{kr_c |y|} \hat{f}_R^{(m) *} \hat{f}_R^{(n)}=\delta^{m n} \ .
    \label{eq:fermionnorm}
\end{align}
The KK modes takes the form, 
\begin{align}
&\hat{f}_{L,R}^{(0)} (y) = \frac{e^{\nu kr_c y}}{N_0^{L,R}} \,, &
\hat{f}_{L, R}^{(n)}(y) = \frac{e^{-kr_c |y| / 2}}{N_n^{L, R}}\left[J_{1 / 2 \mp \nu}\left(z_n^{f}\right)+\beta_n^{L, R} Y_{1 / 2 \mp \nu}\left(z_n^{f}\right)\right] \ ,
\end{align}
where $z_n^f(y)=x_n^f {e^{kr_c (|y|-\pi)}}$ and $\nu = m_q/k$. The normalization constants $N_0^{L,R}$ and $N_n^{L,R}$ are found using eq.\ref{eq:fermionnorm}. The mass of each KK mode is defined as, $$m_n^f = x_n^f ke^{-kr_c\pi} \ ,$$where $x_n^f$ are the solution to the master equation obtain after solving the boundary conditions.
\item {\underline{\it Gauge field}}: To derive the bulk wave-profile of the gauge boson, it suffices to describe a massless $U(1)$ field in the AdS and the generalization to non-Abelian is straight forward. The five dimensional gauge action is given by,
\begin{align}
    S_{\rm gauge} =-\frac{1}{4 g_5^2}\int d^5x \sqrt{-g}\,\left(g^{CM}g^{DN}F_{CD} F_{MN}\right)\,,
\end{align}
where $F_{MN}=\partial_M A_N -\partial_N A_M$ and $g_5$ is the five-dimensional gauge coupling. With the gauge choice $A_4(x_\mu,y)=0$, and assuming the KK expansion of the gauge field as,
\begin{equation}    A_\mu(x_\mu,y)=\sum_{n=0}^{\infty}\frac{1}{\sqrt{\pi r_c}}A_\mu^{(n)}(x_\mu)\chi_A^{(n)}(y) \ ,
\end{equation}
the bulk wave-profiles, upon solving the equation of motion, becomes,
\begin{equation}
    \chi_A^{(n)}=\frac{e^{kr_c |y|}}{N_n^A}\left[J_1\left(z_n^A\right)+\alpha_n^A Y_1\left(z_n^A\right)\right] \ ,
\end{equation}
where $z_n^A(y)=x_n^A {e^{kr_c (|y|-\pi)}}$, and $x_n^A$ are found by solving for the continuity of $d\chi_A^{(n)}/dy$ at $y=0$ and $y=\pi$.

These profiles are set to satisfy the ortho-normality condition,
\begin{equation}
  \int_{-\pi}^\pi d y \chi_A^{(m)} \chi_A^{(n)}=\delta^{m n}  \ .
\end{equation}
For gauge bosons the zero mode wave-profile becomes, $    \chi_A^{(0)} = 1/\sqrt{2\pi}$.
\item {\underline{\it Graviton field}}: The graviton field is a five-dimensional tensor fluctuation about the Minkowski metric given as,
\begin{align}
    G_{\mu \nu}  = e^{-2 k r_c y} (\eta_{\mu \nu} + \kappa_5 h_{\mu \nu}) \ ,
\end{align}
where $\kappa_5= 2 M_5^{-3/2}$. Upon compactification the field can be Fourier decomposed as
\begin{equation}
    h_{\mu\nu}(x_\mu, y)=\sum_{n=0}^{\infty}\frac{1}{\sqrt{\pi r_c}}h_{\mu\nu}^{(n)}(x_\mu)\chi_G^{(n)}(y) \ .
\end{equation}
The $\chi_G^{(n)}(y)$ are set to satisfy the ortho-normality condition,
\begin{equation}
    \int_{-\pi}^\pi d y e^{-2 kr_c |y|} \chi_G^{(m)} \chi_G^{(n)}=\delta^{m n} \ .
\end{equation}
Upon solving the equation of motion, after choosing the gauge $\eta^{\mu \nu}\partial_\mu h^{(n)}_{\nu \gamma}=0$ and $h^{(n) \nu}_{\nu}=0$, we get,
\begin{align}
&\chi_G^{(0)}(y)=\sqrt{k r_c}\,, &\chi_G^{(n)}(y)=\frac{e^{2 kr_c |y|}}{N_n^G}\left[J_2\left(z_n^G\right)+\alpha_n^G Y_2\left(z_n^G\right)\right] \ ,
\end{align}
where $z_n^G=x_n^G {e^{kr_c (|y|-\pi)}}$. The KK excited state masses become, $m_n^G = x_n^G ke^{-kr_c\pi}$ with $x_n^G$ given by the roots of $J_{1}(x_n^G)=0$.
\item {\underline{\it Radion field}}: 
Allowing for small fluctuations about the static solution five-dimensional radius $r_c$, stabilized by the Goldberger-Wise mechanism the metric becomes,
\begin{equation}    ds^2=e^{-2A(y)-2F(x,y)}\,\eta_{\mu\nu}dx^\mu dx^\nu-\left[1+2F(x,y)\right]^2dy^2 \ .
\end{equation}
The lightest solution to the Einstein's equations of motion (called radion), supplemented with the boundary conditions on the field $F$, gives,
\begin{equation}
F_0(x^\mu,y) = e^{2k |y|}(1+ \ell^2 f_0(y)) \frac{1}{\sqrt{6}M_P} e^{-k r_c}r(x^\mu)\,,    
\end{equation}
where $\ell$ is the amount of back-reaction from Goldberger Wise scalar field. While the mass of the radion field becomes,
\begin{equation}
    m^{2}_{\rm rad}= \frac{2\ell^2(2k+u)u^2}{3k} e^{-2(u+k)r_c} \sim \frac{\ell}{40} \text{TeV} \ ,
\end{equation}
where $u$ is the characteristic mass scale obtained in the Goldberger-Wise mechanism. 
For the current discussion we assume that the back-reaction is small ($\ell \ll 1$) making the radion light. 
\end{itemize} 
\subsection{Coupling of bulk fields with KK gravitons}
The couplings in the $5$ dimensions of the $m^{\rm th}$ and the $n^{\rm th}$ KK modes of a general field $F(x,y)$ to $q^{\rm th}$ KK mode of a Graviton ($h_{\mu\nu}(x,y)$) are given by~\cite{Davoudiasl_2001},
\begin{align}
& S= \sum_{m, n, q} \int d^5x \sqrt{g}
\,h_{\mu\nu}^{(q)}(x,y)\,T^{\mu\nu (m,n)}(x,y)   \nonumber\\&
=\sum_{m, n, q}\left\{\left[\int \frac{d y}{\sqrt{k}} \frac{e^{t kr_c |y|} \chi_F^{(m)} \chi_F^{(n)}\,\chi_G^{(q)}}{\sqrt{r_c}}\right] \frac{\kappa_4}{2}\right. \left.\times \int d^4 x\, \eta^{\mu\alpha}\,\eta^{\nu \beta}\,h_{\alpha\beta}^{(q)}(x)\,T_{\mu \nu}^{(m, n)}\right\}\,,
\label{eq:bulkinteractiongraviton}
\end{align} 
where $t$ depends on the type of the field and $\kappa_4/2=1/M_P$. Here, $T_{\mu\nu}$ is the energy momentum tensor of the field under consideration. It is clear from the above expression that the coupling in the $5^{\rm th}$ dimension is given by,
\begin{align}
& \int \frac{d y}{\sqrt{k}} \frac{1}{M_{P}}\frac{e^{t kr_c |y|} \chi_F^{(m)} \chi_F^{(n)} \chi_G^{(q)}}{\sqrt{r_c}}\,.    
\end{align}
The coupling of zeroth modes of any field with the zeroth mode of graviton is equated to $1/M_P$, as all SM couplings in 4D have $1/M_P$ suppression, with
$M_P = 10^{16}\,\text{TeV}$.
Another model dependent parameter is defined as
\begin{align}
\Lambda_{\pi} = \frac{M_P}{8\,\pi} e^{-k\,r_c\,\pi}\,. 
\end{align}
The couplings are calculated in the units of $\Lambda_{\pi}^{-1}$ and normalized by the zeroth mode coupling of the respective field.
\subsubsection{Scalar-graviton coupling}
Scalar-graviton coupling, for a scalar with KK modes localized close to the IR brane, is obtained using,
\begin{align}
C_{mnq}^{DDG}=\int_{-\pi}^\pi \frac{d y}{\sqrt{k}}\frac{1}{M_{P}} \frac{e^{-2 kr_c |y|}\chi_D^{(m)} \chi_D^{(n)} \chi_G^{(q)}}{\sqrt{r_c}}\ .
\end{align} 
For a choice of the bulk mass parameter, where the scalar is localized towards to UV, the relevant couplings are reported in Tab.~\ref{tab:scalar_graviton}.
\begin{table}[h]
\centering
\begin{tabular}{|c|c|c|c|c|c|c|}
\hline
\multicolumn{2}{|c|}{Scalar Field} & \multicolumn{2}{c|}{Scalar Field} & \multicolumn{2}{c|}{Graviton} & Coupling\\
\hline
$m$ & $m_m = x_m\cdot\Lambda_{\pi}$ & $n$ & $m_n = x_n\cdot\Lambda_{\pi}$ & $q$ & $m_q = x_q\cdot\Lambda_{\pi}$ & $\times \Lambda_{\pi}^{-1}$ \\ \hline
\hline
0 & 0    & 0 & 0    & 1 & 3.83  &  \(1.03298 \times 10^{-15}\) \\
\hline
0 & 0    & 0 & 0    & 2 & 7.02  &  \(5.70229 \times 10^{-16}\) \\
\hline
0 & 0    & 0 & 0    & 3 & 10.17 &  \(4.89726 \times 10^{-16}\) \\
\hline
0 & 0    & 0 & 0    & 4 & 13.32 &  \(3.84256 \times 10^{-16}\) \\
\hline
1 & 3.01 & 1 & 3.01 & 1 & 3.83  &  0.0224298 \\
\hline
1 & 3.01 & 1 & 3.01 & 2 & 7.02  &  0.0080403 \\
\hline
1 & 3.01 & 1 & 3.01 & 3 & 10.17 &  0.00172851 \\
\hline
1 & 3.01 & 1 & 3.01 & 4 & 13.32 &  0.00059826 \\
\hline
\end{tabular}
\caption{Masses and coupling values for scalar-graviton interactions.}
\label{tab:scalar_graviton}
\end{table}
\subsubsection{Fermion-graviton coupling}
These couplings are given by the integral~\cite{Davoudiasl_2001},
\begin{align}
C_{m n q}^{ffG}=\int_{-\pi}^\pi \frac{d y}{\sqrt{k}}\frac{1}{M_{P}} \frac{e^{ kr_c |y|} \hat{f}_L^{(m)} \hat{f}_L^{(n)} \chi_G^{(q)}}{\sqrt{r_c}}\,,    
\end{align}
and are tabulated in Tab.~\ref{tab:fermion_graviton}.
Throughout the calculation the bulk parameter such that the fermion is localized towards the IR brane. The femrion couplings are with the heavier KK modes of graviton is given in Tab. \ref{tab:fermion_graviton}.
\begin{table}[h]
\centering
\begin{tabular}{|c|c|c|c|c|c|c|}
\hline
\multicolumn{2}{|c|}{Fermion Field} & \multicolumn{2}{c|}{Fermion Field} & \multicolumn{2}{c|}{Graviton Field} & Coupling  \\ \hline
$m$ & $m_m = x_m\cdot\Lambda_{\pi}$ & $n$ & $m_n = x_n\cdot\Lambda_{\pi}$ & $q$ & $m_q = x_q\cdot\Lambda_{\pi}$ & $\times \Lambda_{\pi}^{-1}$ \\ \hline
0 & 0 & 0 & 0 & 1 & 3.83 & 0.0184379 \\ \hline
0 & 0 & 0 & 0 & 2 & 7.02 & 0.00392329 \\ \hline
0 & 0 & 0 & 0 & 3 & 10.17 & 0.0022257 \\ \hline
0 & 0 & 0 & 0 & 4 & 13.32 & 0.00111986 \\ \hline
1 & 3.51 & 1 & 3.51 & 1 & 3.83 & 0.0146376 \\ \hline
1 & 3.51 & 1 & 3.51 & 2 & 7.02 & 0.00846008 \\ \hline
1 & 3.51 & 1 & 3.51 & 3 & 10.17 & 0.00809693 \\ \hline
1 & 3.51 & 1 & 3.51 & 4 & 13.32 & 0.00180826 \\ \hline
\end{tabular}
\caption{Masses and couplings relevant for fermion-graviton interaction.}
\label{tab:fermion_graviton}
\end{table}

\subsubsection{Gauge-graviton coupling}
The coupling between gauge fields and graviton are be obtained from the integral~\cite{Davoudiasl_2001},
\begin{align}
C_{m n q}^{VVG}=\int_{-\pi}^\pi \frac{d y}{\sqrt{k}}\frac{1}{M_{P}} \frac{\chi_A^{(m)} \chi_A^{(n)} \chi_G^{(q)}}{\sqrt{r_c}}\,,    
\end{align}
where $A$ denotes the gauge boson. Their numerical values are tabulated in Tab.~\ref{tab:gauge_graviton}.
\begin{table}[h]
\centering
\begin{tabular}{|c|c|c|c|c|c|c|}
\hline
\multicolumn{2}{|c|}{Gauge Field} & \multicolumn{2}{c|}{Gauge Field} & \multicolumn{2}{c|}{Graviton Field} & Coupling   \\ \hline
$m$ & $m_m = x_m\cdot\Lambda_{\pi}$ & $n$ & $m_n = x_n\cdot\Lambda_{\pi}$ & $q$ & $m_q = x_q\cdot\Lambda_{\pi}$ & $\times \Lambda_{\pi}^{-1}$ \\ \hline
0 & 0 & 0 & 0 & 1 & 3.83 & 0.00214982 \\ \hline
0 & 0 & 0 & 0 & 2 & 7.02 & 0.00042603 \\ \hline
0 & 0 & 0 & 0 & 3 & 10.17 & 0.000436992 \\ \hline
0 & 0 & 0 & 0 & 4 & 13.32 & 0.000184107 \\ \hline
1 & 2.45 & 1 & 2.45 & 1 & 3.83 & 0.0983561 \\ \hline
1 & 2.45 & 1 & 2.45 & 2 & 7.02 & 0.0315747 \\ \hline
1 & 2.45 & 1 & 2.45 & 3 & 10.17 & 0.00713964 \\ \hline
1 & 2.45 & 1 & 2.45 & 4 & 13.32 & 0.00350777 \\ \hline
\end{tabular}
\caption{Masses and couplings relevant for the gauge-graviton interaction.}
\label{tab:gauge_graviton}
\end{table}

\subsection{Radion coupling with fields}
The radion interaction is given by,
\begin{align}
\mathcal{L}_{\rm radion}=\int_{-\pi}^{\pi} dy F_0(x^\mu,y)\,{\rm Tr}(T_{\mu\nu} (x^\mu,y)) = C^r r(x^\mu)\,{\rm Tr}(T_{\mu\nu}(x^\mu))\,,    
\end{align}
where, for IR localized energy momentum tensor,
\begin{align}
C^r = \int_{-\pi}^{\pi}\,dy\,e^{2k |y|}\frac{1}{\sqrt{6}M_P}\,e^{-k r_c}\,\delta(y-r_c) = \frac{1}{\Lambda_W} \ ,
\end{align}
where $\Lambda_W \sim 1 \text{ TeV}$, whereas for UV localized matter,
\begin{align}
C^r = \int_{-\pi}^{\pi}dy\, e^{2k |y|}\,\frac{1}{\sqrt{6}M_P}\,e^{-k r_c}\,\delta(y) = \frac{v}{\sqrt{6} M_P^2}\,.
\end{align}
Although the radion couples to IR-localized fields in a manner similar to the graviton, its couplings to UV-localized fields are suppressed by several orders of magnitude relative to those of the graviton, as opposed to the scenario discussed in~\cite{Bernal:2020fvw}. For bulk fermions, a numerical analysis confirms that, for the range of bulk mass parameters considered in this work, radion-mediated interactions are significantly weaker than their graviton-mediated counterparts. Furthermore, the radion mass depends on the extent of backreaction on the geometry and typically lies within the range $[1~\text{eV} : 10~\text{GeV}]$. As a result, at high center-of-mass energies, radion-mediated cross-sections grow as $\sim s$, whereas graviton-mediated ones scale as $\sim s^3$. Since DM production in our scenario is dominated by UV-scale dynamics, graviton-mediated processes overwhelmingly dominate over those mediated by the radion. Therefore, we neglect contributions from radion exchange diagrams in this analysis.

Before closing this section, we briefly discuss phenomenological bounds on warped extra dimensional scenario arising from collider searches and measurement of precision observables. The primary decay channel for the KK graviton is through a pair of top quarks~\cite{Davoudiasl:1999jd, Davoudiasl:2009cd}. Direct searches for KK gravitons at CMS with $\sqrt{s}= 4$TeV and integrated luminosity of $4$ fb${}^{-1}$~\cite{CMS:2013egk}, in the dijet channel at LHC with QCD NLO accuracy, exclude the parameter space below $M_{\rm KK}=1.45$ TeV~\cite{Li:2014awa}. While, including the decay of the final state electroweak gauge bosons to di-lepton and di-photon channel with $\sqrt{s}=14$ TeV and $50$ fb${}^{-1}$ integrated luminosity, places a very restrictive limit of $5.2$ TeV~\cite{Das:2014tva}. Though this might seem strict, it is model dependent and will be relaxed if we assume the light quarks to be localized further to the UV brane, essentially reducing the production cross-section of the KK graviton with minimal effect to the anarchic Yukawa explaining the fermion mass hierarchy. Realistic models, assuming custodial protection~\cite{Agashe:2003zs}, and protection of $Z b_L\,b_L$~\cite{Agashe:2006at} coupling, keeps the lower limit on the masses of the KK partners of the electroweak gauge bosons to $M_{\rm KK} > 3$ TeV safe from electro-weak precision tests. While in the flavour violations, the RS-GIM mechanism~\cite{Huber:2003tu,Agashe:2004cp} along with the Minimal Flavour Protection paradigm~\cite{Santiago:2008vq} is incredible in suppressing the large flavour violating currents in the right handed down sector mediated by the KK towers of gluons. Summarizing, a scale of 3\----4 TeV is compatible with both electroweak precision observable (EWPO) and flavour physics.
\section{Freeze-in production of dark matter}
\label{sec:dm}
In this section we discuss the DM genesis mechanism. The DM is a feebly interacting massive particle produced via freeze-in\footnote{The possibility of freeze-out has been studied in the framework of RS scenario in~\cite{Lee:2013bua,Lee:2014caa,Rueter:2017nbk,Rizzo:2018joy,Rizzo:2018ntg,Carrillo-Monteverde:2018phy,Brax:2019koq,Folgado:2019sgz}.}. As a result, it never reaches thermal equilibrium with the SM thermal bath, and hence its abundance remains smaller than the equilibrium one along the history of the Universe. The key assumption of freeze-in is to consider a null and void dark sector at an early epoch. The dark sector gets populated with DM, that are produced form the visible sector via some DM-SM interaction, as the Universe evolves with time. Here we typically concentrate on DM production from the scattering of the SM particles and their KK-modes in the initial states, mediated by massive gravitons. We take up three different DM spins, namely, spin-0 scalar, spin-1/2 (Dirac) fermion and spin-1 massive vector boson, and consider each of them to be present one at a time. We begin with the discussion on the dark brane composite DM production and then we move on to the  production of partially composite DM production. As we will show, the DM production channels are distinctively different in the two scenarios. 
\subsection{Dark brane composite DM production}
In order to track the evolution of DM number density $n_{\rm DM}$ with time, we write down the Boltzmann equation (BEQ) as follows,
\begin{align}
\dot n_{\rm DM}+3\,H\,n_{\rm DM}=\gamma_{22}\,,    
\end{align}
where the reaction density in case of 2-to-2 scattering is given by~\cite{Duch:2017khv}
\begin{align}
& \gamma_{22}=\frac{T}{32\pi^4}\,g_a g_b
\nonumber\\&
\times\int_{\text{max}\left[\left(m_a+m_b\right)^2,\left(m_1+m_2\right)^2\right]}^\infty ds\,\frac{\biggl[\bigl(s-m_a^2-m_b^2\bigr)^2-4m_a^2 m_b^2\biggr]}{\sqrt{s}}\,\sigma\left(s\right)_{a,b\to1,2}\,K_1\left(\frac{\sqrt{s}}{T}\right)\label{eq:gam-ann}\,. \end{align}    
Here $a,b\,(1,2)$ are the incoming (outgoing) states and $g_{a,b}$ are the corresponding internal degrees of freedom of the initial states. In the present scenario $a,\,b$ are the SM states or their KK-partners, as shown in Fig.~\ref{fig:feyn}.  In a radiation-dominated Universe, where the entropy per comoving volume is conserved, the above equation can be re-casted in terms of the DM yield $Y_{\rm DM}=n_{\rm DM}/s$ as\footnote{We are considering direct freeze-in production of the DM, where both the SM and the KK-partners are considered to be in equilibrium with the bath. This can be justified by noting that the KK-partners also have SM gauge-boson mediated interactions that are not suppressed by the Planck scale, while there is no direct coupling with the DM.}
\begin{equation}\label{eq:beq}
x\,H\,\mathfrak{s}\,\frac{dY_{\rm DM}}{dx} =\gamma_{22}(T)\,,
\end{equation}
where $x\equiv\mdm/T$ is a dimensionless quantity with $T$ being the temperature of the thermal bath. For a radiation dominated Universe the entropy density $s$ and Hubble parameter $H$ are given by, 
\begin{align}
& \mathfrak{s}(T)=\frac{2\,\pi^2}{45}\,\gss(T)\,T^3\,, & 
H(T)=\frac{\pi}{3}\,\sqrt{\frac{\gs(T)}{10}}\,\frac{T^2}{M_P}\,,
\end{align}
where $\gss$ and $\gs$ are the effective number of relativistic degrees of freedom contributing to the entropy and energy density respectively. The DM yield at a temperature $T$ cab be obtained by integrating Eq.~\eqref{eq:beq} as,
\begin{align}\label{eq:dm-yield}
& Y_{\rm DM}(T)=-M_P\,\int_{\Trh}^T\,\mathcal{C}(T)\,\frac{\gamma_{\rm tot}(T)}{T^6}\,,    
\end{align}
where $\mathcal{C}(T)=\left(2\pi^2/45\right)\,\gss(T)\,\sqrt{\pi^2\,\gs(T)/90}$ and $\gamma_{\rm tot}$ represents the total reaction density, including all KK-modes and processes. Here $\Trh$ is the reheating temperature which, in the approximation of a sudden decay of the inflaton, corresponds to the maximal temperature reached by the SM thermal bath. In obtaining Eq.~\eqref{eq:dm-yield} a vanishing initial DM abundance at $T=\Trh$ is assumed, which holds true for freeze-in. To fit the whole observed DM relic density, it is required that
\begin{equation} \label{eq:obsyield}
    Y_0\, \mdm = \Omega h^2 \, \frac{1}{s_0}\,\frac{\rho_c}{h^2} \simeq \mathcal{C}_{\rm rel}\,,
\end{equation}
where $Y_0 \equiv Y_{\rm DM}(T_0)$ is the DM yield today. Furthermore, $\rho_c \simeq 1.05 \times 10^{-5}\, h^2$~GeV/cm$^3$ is the critical energy density, $s_0\simeq 2.69 \times 10^3$~cm$^{-3}$ the present entropy density~\cite{ParticleDataGroup:2022pth}, and $\Omega h^2 \simeq 0.12$ the observed abundance of DM relics~\cite{Planck:2018vyg}, with $\mathcal{C}_{\rm rel}=4.3 \times 10^{-10}$ GeV. 
\begin{figure}
    \centering         \includegraphics[scale=0.13]{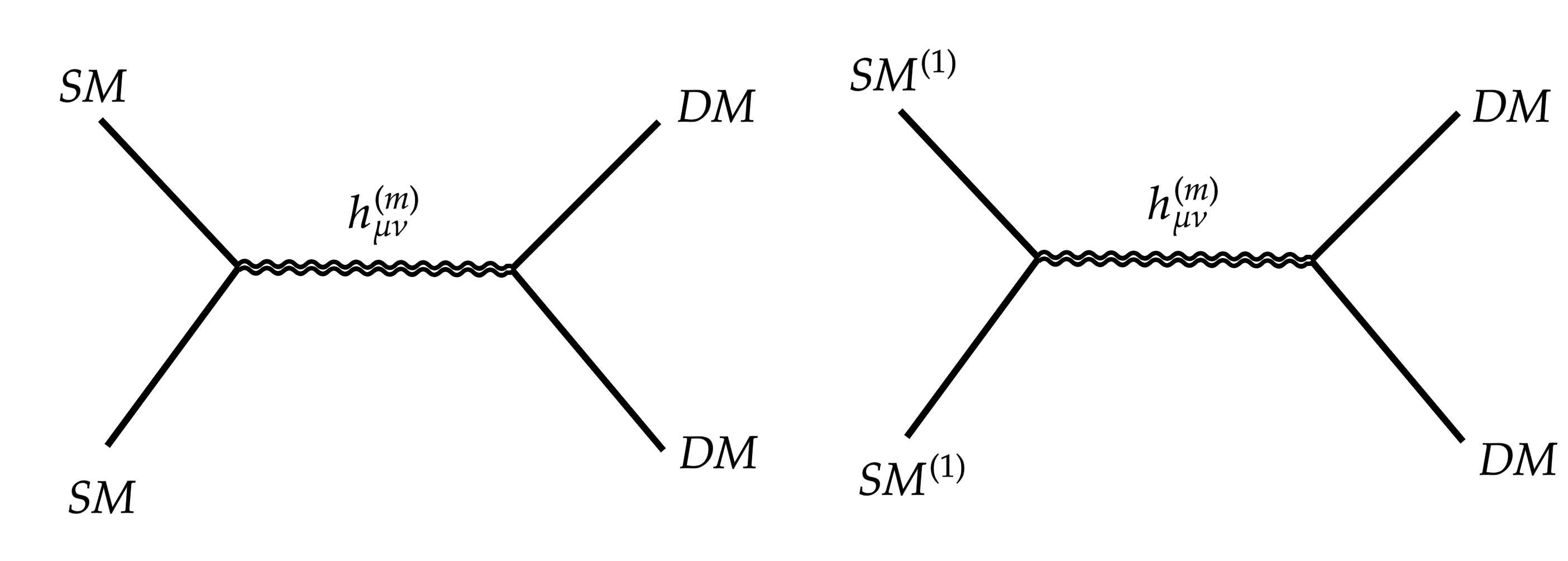}   \caption{Annihilation of SM and their KK partners into a pair of DM, mediated by massive graviton propagators denoted by $h_{\mu\nu}$. Here `DM' stands for either a spin-0, a spin-1/2 or a spin-1 field.}
    \label{fig:feyn}
\end{figure}
\begin{figure}[htb!]
    \centering
    \includegraphics[scale=0.375]{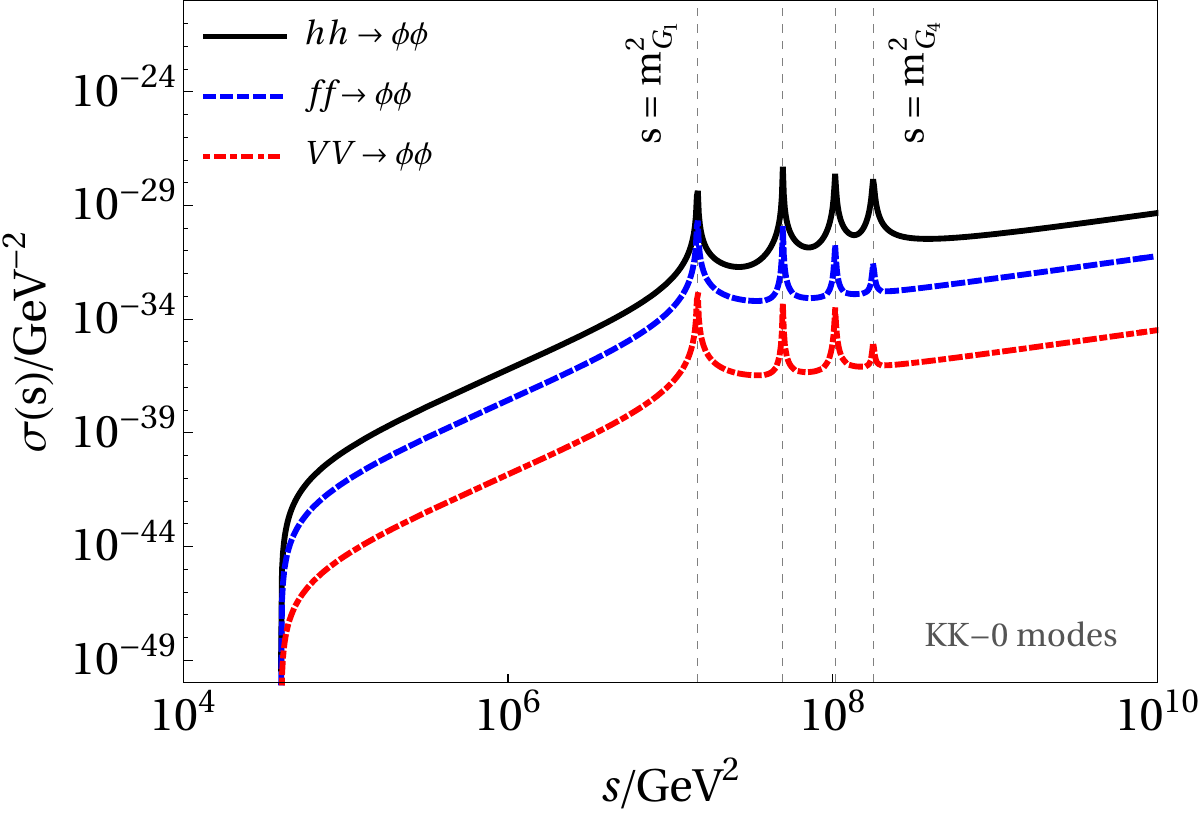}~\includegraphics[scale=0.375]{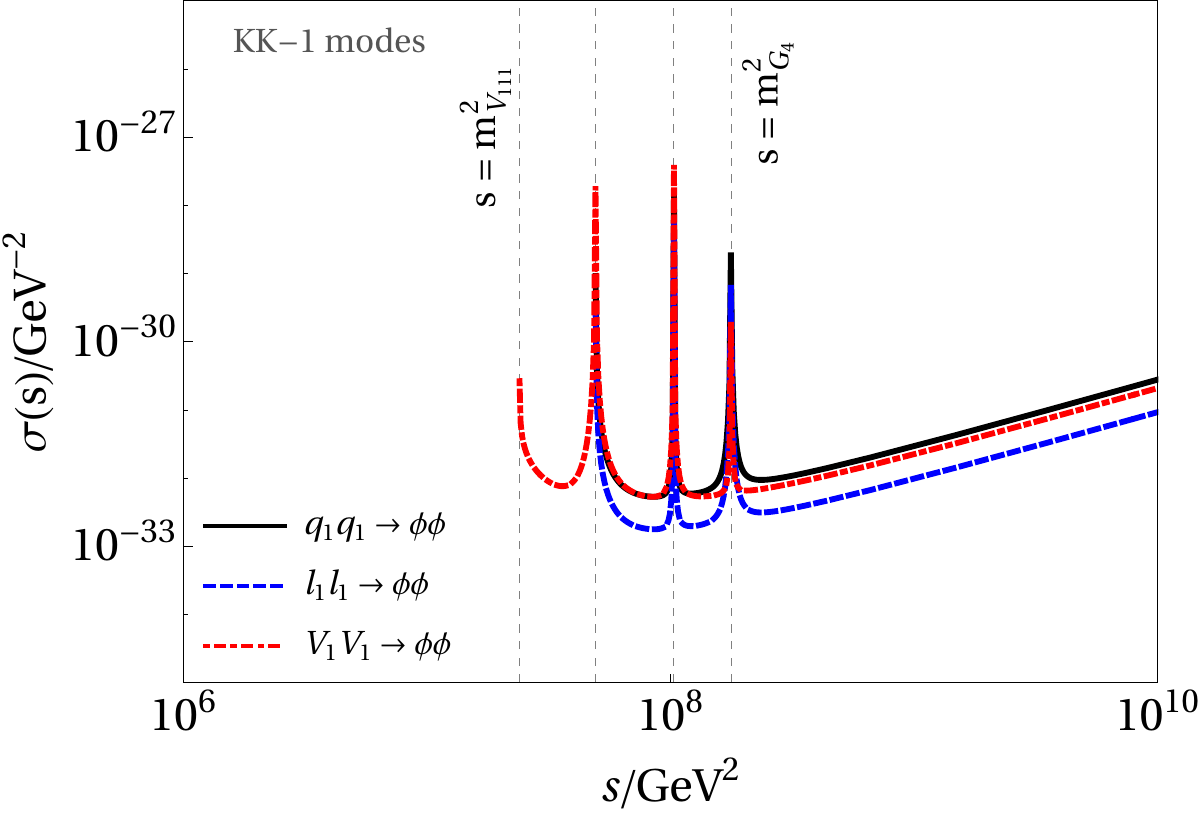}\\[10pt]
    \includegraphics[scale=0.5]{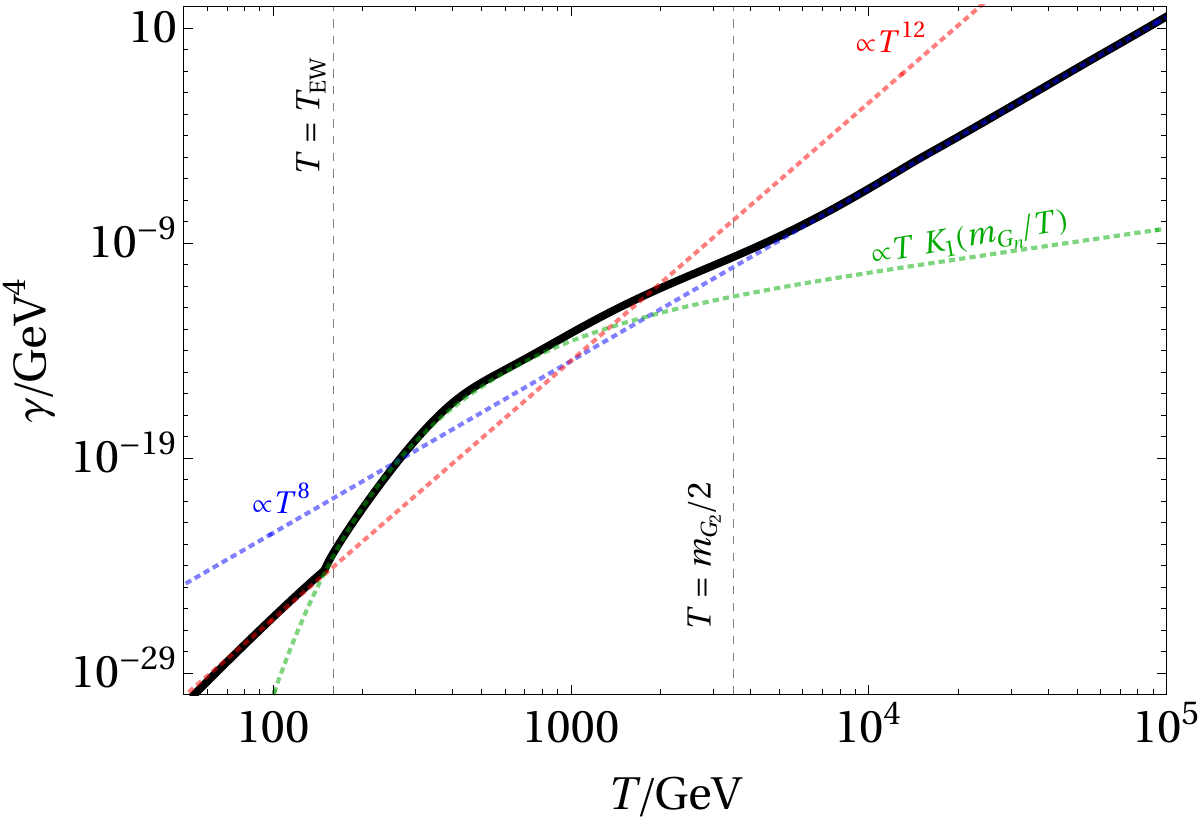}
    \caption{Cross-section into spin-0 DM final states mediated by massive gravitons, as a function of $s$, considering only the zero modes (top left pane) and first modes (top right panel). In the bottom panel we show the total reaction density, taking all KK modes into account. The dotted lines correspond to approximate analytical estimation, obtained in Eq.~\eqref{eq:gamma-analyt}. Here we have fixed $\mdm=100$ GeV, $\Lambda_\pi=1$ TeV and $C_{000}^{\phi\phi G}=10^{-15}\,\text{GeV}^{-1}$.}
    \label{fig:cs}
\end{figure}

In the top panel fo Fig.~\ref{fig:cs} we show the variation of annihilation cross-sections into DM final states (following Fig.~\ref{fig:feyn}), as a function of the squared of the center of mass energy scale $\sqrt{s}$. For illustrative purposes, here we have considered the scalar DM case. The full expression for all relevant cross-sections and decay rates are reported in Appendix.~\ref{sec:cs}. For both zero and first KK-modes, we note resonant enhancements in the cross-section corresponding to $s=m_{G_n}^2$. In all cases we have fixed DM mass to 100 GeV. As a consequence, for annihilation of the KK-0 modes (top left panel), we see a cut-off at $s=4\,\mdm^2$. This is because, in this case, the initial state particles (Higgs and top-quark) have masses comparable to that of the DM mass after the electroweak symmetry (EW) is broken, while before that they are considered to be absolutely massless. For annihilation of the first KK-modes, as shown in the top right panel, the cut-off appears for $s=4\,m_{V_{111}}^2$. As the cross-sections scale as $\sigma(s)\sim s^3$, hence, in all cases they rise with $s$. It is worth noting that the contribution from the $q=4$ KK mode is reduced compared to that from the $q=2$ mode, irrespective of the initial state. Thus, contributions from KK modes $q\gg 4$ are expected to be even more suppressed compared to the $q=2$ mode. Before moving on let us clarify that, following are the set of independent parameters in the present framework, that we will constraint via DM phenomenology:
\begin{align}
\Big\{\Lambda_\pi,\,\mdm,\,\Trh,\,C_{mnq}^{\phi\phi G}\Big\}\,.    
\end{align}

In the bottom panel of Fig.~\ref{fig:cs} we illustrate the evolution of the reaction density $\gamma(T)$, as a function of the bath temperature, where we have included all the KK-modes. In order to physically understand the features appearing in this figure, we obtain an approximate analytical expression of the reaction density. This can be done in the limit where the initial and final states have negligible masses with respect to $\sqrt{s}$. In the  case of spin-0 DM, the total cross-section can be approximately written as
\begin{align}
& \sigma(s)\simeq \frac{s^3\,\left(C_{mnq}^{\phi\phi G}\right)^2}{8640\,\pi}\,\left[\frac{3}{\Lambda_\pi^2}+ \left(\left(3\,C_{mnq}^{ffG}\right)^2+\left(C_{mnq}^{VVG}\right)^2\right)\right]\,
\Bigg|\sum_{n=1}^\infty\,\frac{1}{s-m_{G_n}^2+i\,\Gamma_{G_n}\,m_{G_n}}\Bigg|^2\,.  
\end{align}
In this case, the reaction density simplifies to
\begin{align}
&\gamma(T)\simeq\frac{T}{8\,\pi^4}\,\int_{4\,\mdm^2}^\infty\,ds\,s^{3/2}\,\sigma(s)\,K_1\left(\frac{\sqrt{s}}{T}\right)\,. 
\end{align}
Now, near the resonance, the so-called narrow width approximation
\begin{equation}
    \frac{1}{(s-m_{G_n}^2)^2 + m_{G_n}^2\, \Gamma^2} \to \frac{\pi}{m_{G_n}\, \Gamma}\, \delta\left(s - m_{G_n}^2\right)\,,
\end{equation}
is valid, while away from the resonance, one may ignore the graviton decay width, and consider the propagator to be $1/(s-m_{G_n}^2)^2$. It is then possible to obtain an approximate analytical expression for the reaction density as
\begin{align}
& \gamma(T)\simeq\frac{\left(C_{mnq}^{\phi\phi G}\right)^2}{\pi^4}\,\left[\frac{3}{\Lambda_\pi^2}+ \left(3\,C_{mnq}^{ffG}\right)^2+\left(C_{mnq}^{VVG}\right)^2\right]
\nonumber\\&
\times
\begin{dcases}
\frac{128\,T^{12}}{3\,\pi\,m_{G_n}^4}\,, & T\ll m_{G_n}/2\,,
\\[10pt]
\frac{m_{G_n}^8\,T}{69120\,\Gamma_{G_n}}\,K_1\left(\frac{m_{G_n}}{T}\right)\,, & T\simeq m_{G_n}/2\,,
\\[10pt]
\frac{T^8}{90\,\pi^5}\,, & T\gg m_{G_n}/2\,,
\end{dcases}
\label{eq:gamma-analyt}
\end{align}
where $K_1[x]$  denotes the modified Bessel function of the first kind. We now clearly see that, at low temperatures $T\ll m_{G_n}/2$, all the mediators are very heavy and decouple from the low-energy theory, and the rate presents a strong temperature dependence $\gamma\propto T^{12}$ (shown by red dotted line). For $T\simeq m_{G_n}/2$, the resonant exchange of graviton dominates leading to $\gamma\propto T\,K_1\left(m_{G_n}/T\right)$ (shown by the green dotted curve). This corresponds to the ``bump" in the curve. Finally, for $T\gg m_{G_n}/2$, the mediator mass is negligible, resulting in $\gamma\propto T^8$ (shown by the blue dotted line). Thus, the approximate behavior obtained analytically in Eq.~\eqref{eq:gamma-analyt} exactly reproduces the scaling of $\gamma(T)$ with temperature, obtained numerically in the bottom panel of Fig.~\ref{fig:cs}. 

Using Eq.~\eqref{eq:gamma-analyt}, we can perform the integration in Eq.~\eqref{eq:dm-yield} analytically, and obtain the final DM yield as
\begin{align}
& Y_0\simeq\mathcal{C}(T)\,M_P\,\left(C_{mnq}^{\phi\phi G}\right)^2\,\left[\frac{3}{\Lambda_\pi^2}+ \left(3\,C_{mnq}^{ffG}\right)^2+\left(C_{mnq}^{VVG}\right)^2\right]
\nonumber\\&
\times
\begin{dcases}
\frac{128}{21\,\pi^5}\,\frac{\Trh^7}{m_{G_n}^4}\,, & \Trh\ll m_{G_n}/2\,,
\\[10pt]
\frac{m_{G_n}^{9/2}\,e^{-m_{G_n}/\Trh}}{276480\,\pi^4\,\Trh^{5/2}\,\Gamma_{G_n}}\,\left(4\,m_{G_n}^2+10\,m_{G_n}\,\Trh+15\,\Trh^2\right)\,, & \Trh\simeq m_{G_n}/2\,,
\\[10pt]
\frac{\Trh^3}{270\,\pi^5}\,, & \Trh\gg m_{G_n}/2\,,
\end{dcases}
\label{eq:yld-analyt}
\end{align}
where in the second line we have used an approximate expansion of the Bessel function as, $K_1[z]\simeq z^{-1/2}\,e^{-z}$. The above equation shows that the asymptotic DM yield has a strong dependence on the reheating temperature, a quintessential feature of UVFI~\cite{Hall:2009bx,Elahi:2014fsa}. 
\begin{figure}[htb!]    
\centering
\includegraphics[scale=0.375]{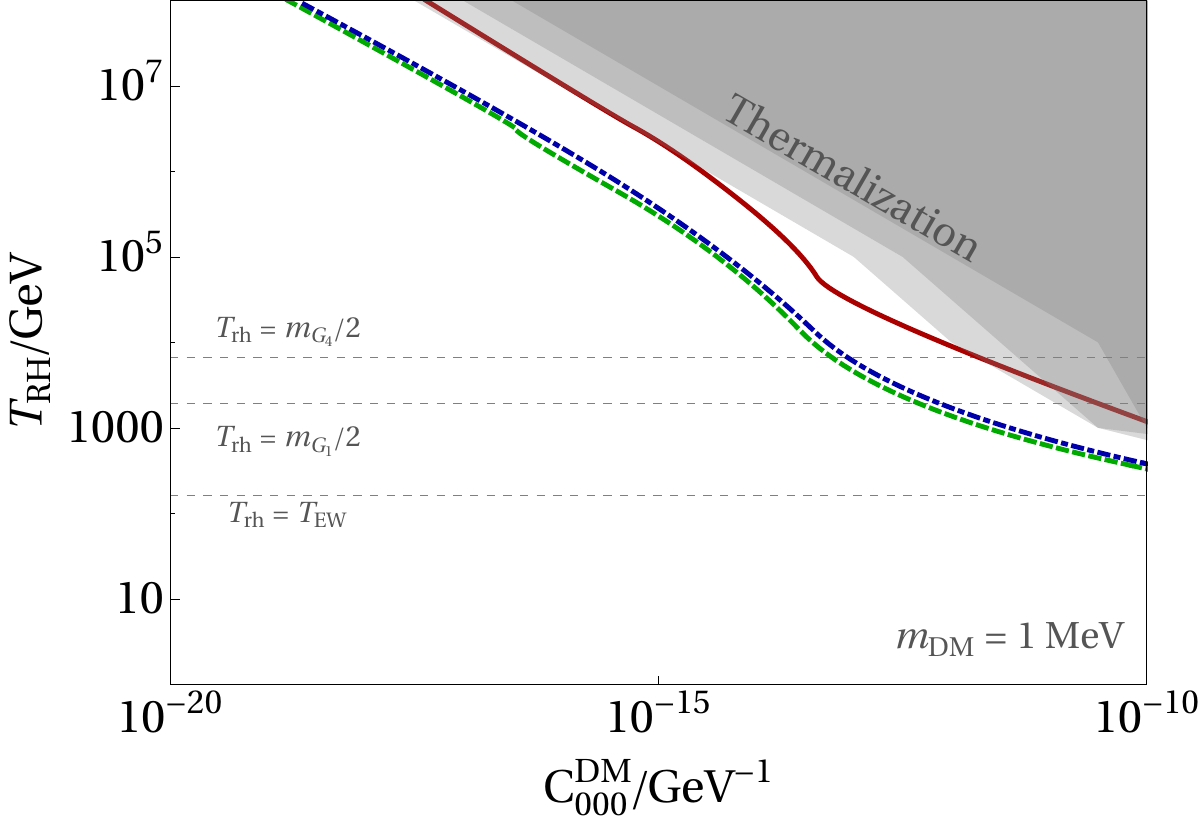}~~
\includegraphics[scale=0.375]{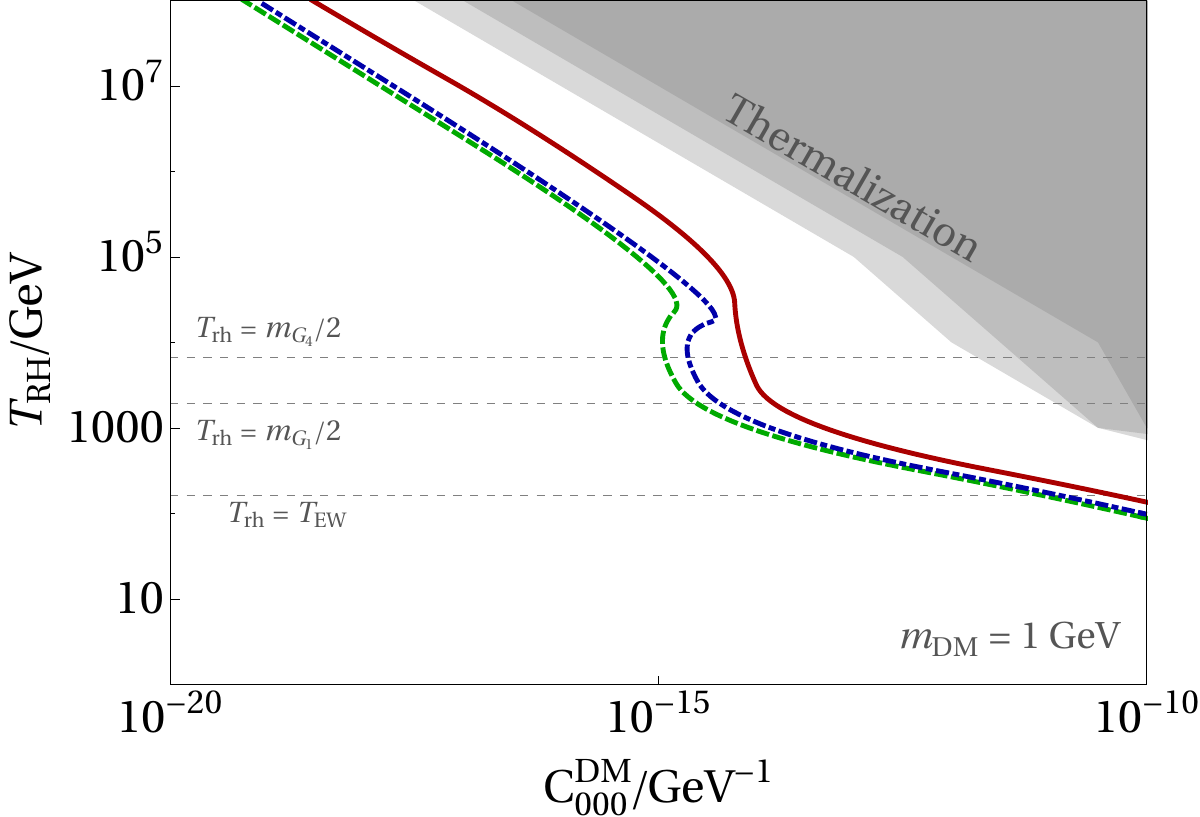}
\\[10pt]
\includegraphics[scale=0.375]{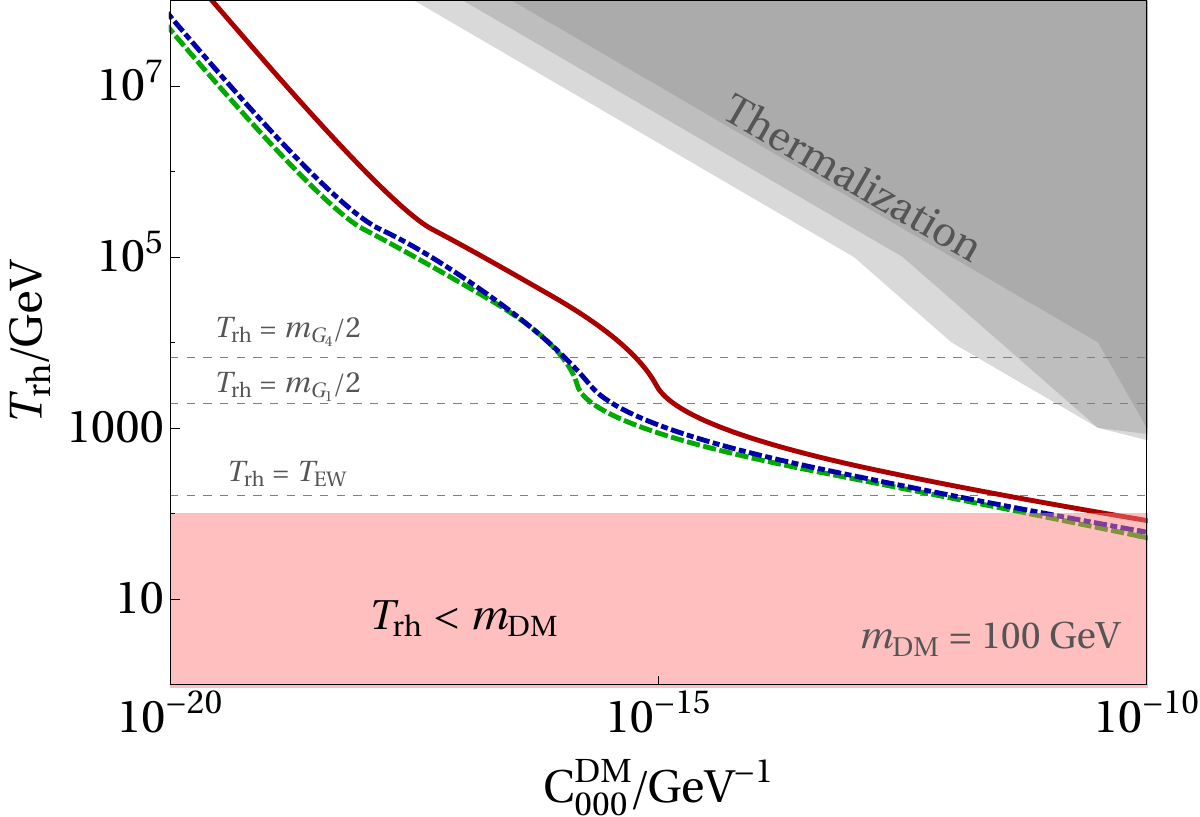}~\includegraphics[scale=0.375]{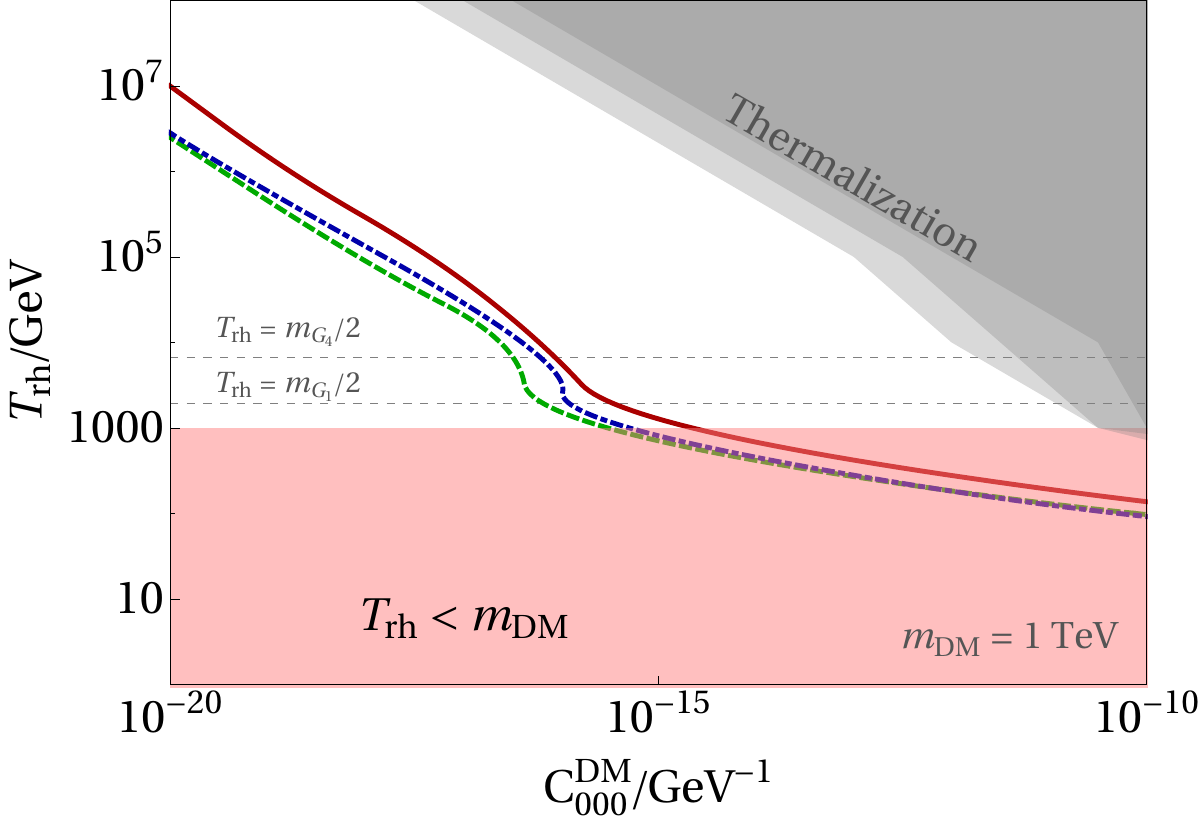}
\caption{Contours of observed DM abundance for a spin-0 (solid, red), spin-1/2 (dashed, green) and spin-1 (dot-dashed, blue) DM of mass 1 MeV (top left), 1 GeV (top right), 100 GeV (bottom left) and 1 TeV (bottom right). The gray shaded region is forbidden from the condition of thermalization (see text). The lighter to darker shade corresponds to spin-1/2, spin-1 and spin-0, respectively. In the bottom panel the red shaded region is forbidden due to $\mdm>\Trh$. Here we have fixed $\Lambda_\pi=1$ TeV.}
\label{fig:relic}
\end{figure}

For the previous analysis to be valid, the DM has to be out of chemical equilibrium with the radiation bath, which is the primary requirement for freeze-in. One needs to therefore guarantee that the
interaction rate density should satisfy 
\begin{align}\label{eq:dm-therm}
\gamma(\Trh)<n_{\rm eq}\,H(\Trh)\,,  
\end{align}
since at $T=\Trh$ bulk of the DM is produced. Considering $n_{\rm eq}^{\rm DM}(T)=\frac{T}{2\pi^2}\,\mdm^2\,K_1\left(\frac{\mdm}{T}\right)$, we obtain the following limits on $\Trh$,
\begin{align}
& \Trh<
\begin{dcases}
\left[\frac{\Lambda_\pi^2\,m_{G_n}^4}{2\,M_P\,\left(C_{000}^{\phi\phi G}\right)^2}\right]^{1/7}\,\left(\frac{\sqrt{\gs}\,\pi^4}{\Upsilon}\right)^{1/7}\,, & \Trh\ll m_{G_n}/2\,,
\\[10pt]
\frac{-0.28\,m_{G_n}}{\mathcal{W}_0\,\left[-0.34\,\left(\frac{m_{G_n}^4}{\mathcal{A}\,\mdm^2}\right)^{2/7}\right]}\,, & \Trh\simeq m_{G_n}/2\,,
\\[10pt]
\left[\frac{\Lambda_\pi^2}{M_P\,\left(C_{000}^{\phi\phi G}\right)^2}\right]^{1/3}\,\left(\frac{30\,\pi^4\,\sqrt{\gs}}{\Upsilon}\right)^{1/3}\,, & \Trh\gg m_{G_n}/2\,,
\end{dcases}
\label{eq:Trh-analyt}
\end{align}
where $\mathcal{W}_0[x]$ is the principal branch of the Lambert $\mathcal{W}$ function, and $$\mathcal{A}=\left(\frac{\Upsilon\,\left(C_{000}^{\phi\phi G}\right)^2\,m_{G_n}^8\,M_P}{11520\,\pi^3\,\sqrt{\gs}\,\mdm^2\,\Gamma_{G_n}\,\Lambda_\pi^2}\right)\,\text{with}\,\Upsilon=3+ 16\,\Lambda_\pi^2\,\left[\left(3\,C_{mnq}^{ffG}\right)^2+\left(C_{mnq}^{VVG}\right)^2\right]\,.$$ In deriving the above expressions we have considered an equilibrium DM number density, which provides rather a conservative bound on the reheating temperature. We emphasize that although Eqs.~\eqref{eq:gamma-analyt}, \eqref{eq:yld-analyt} and \eqref{eq:Trh-analyt} are written considering the scalar DM scenario, for spin-1/2 and spin-1 cases we obtain similar dependence on mass and temperature.
\begin{figure}[htb!]
    \centering    \includegraphics[scale=0.375]{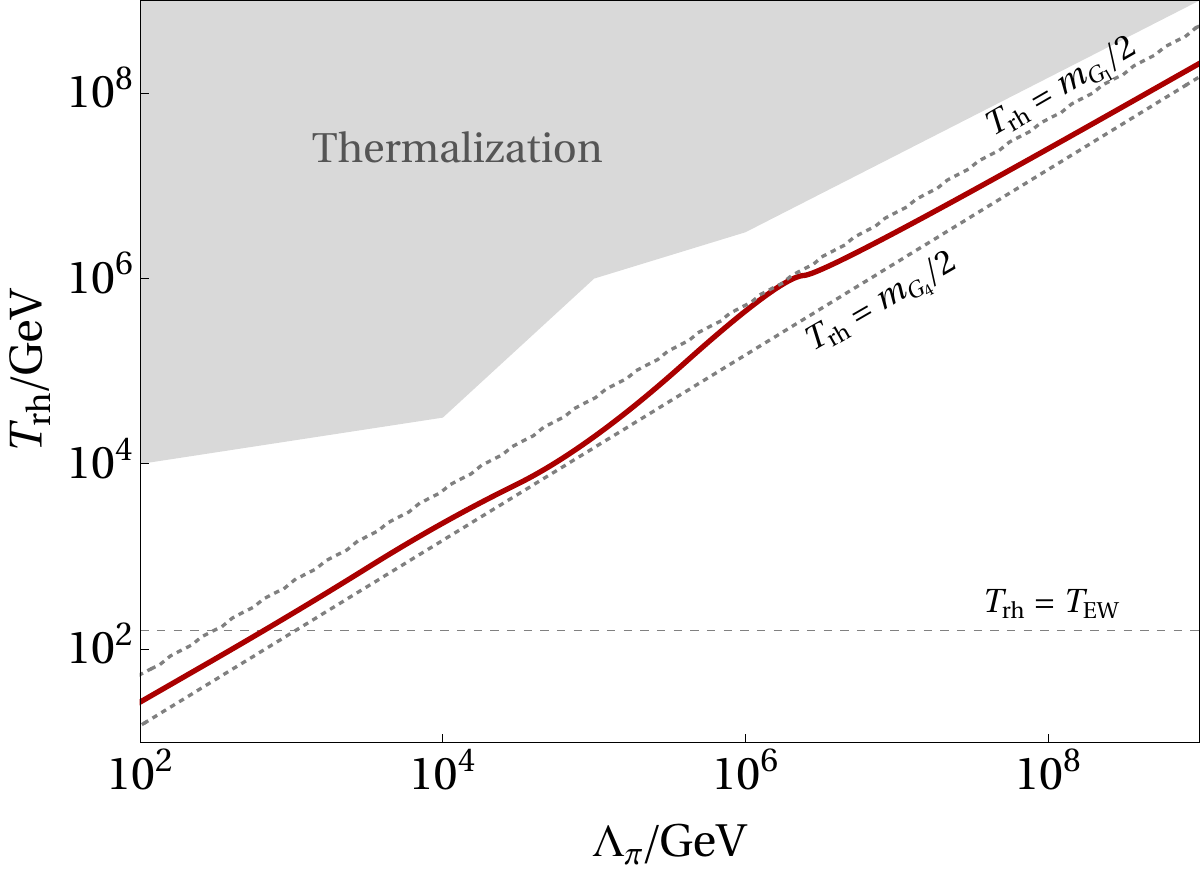}~\includegraphics[scale=0.375]{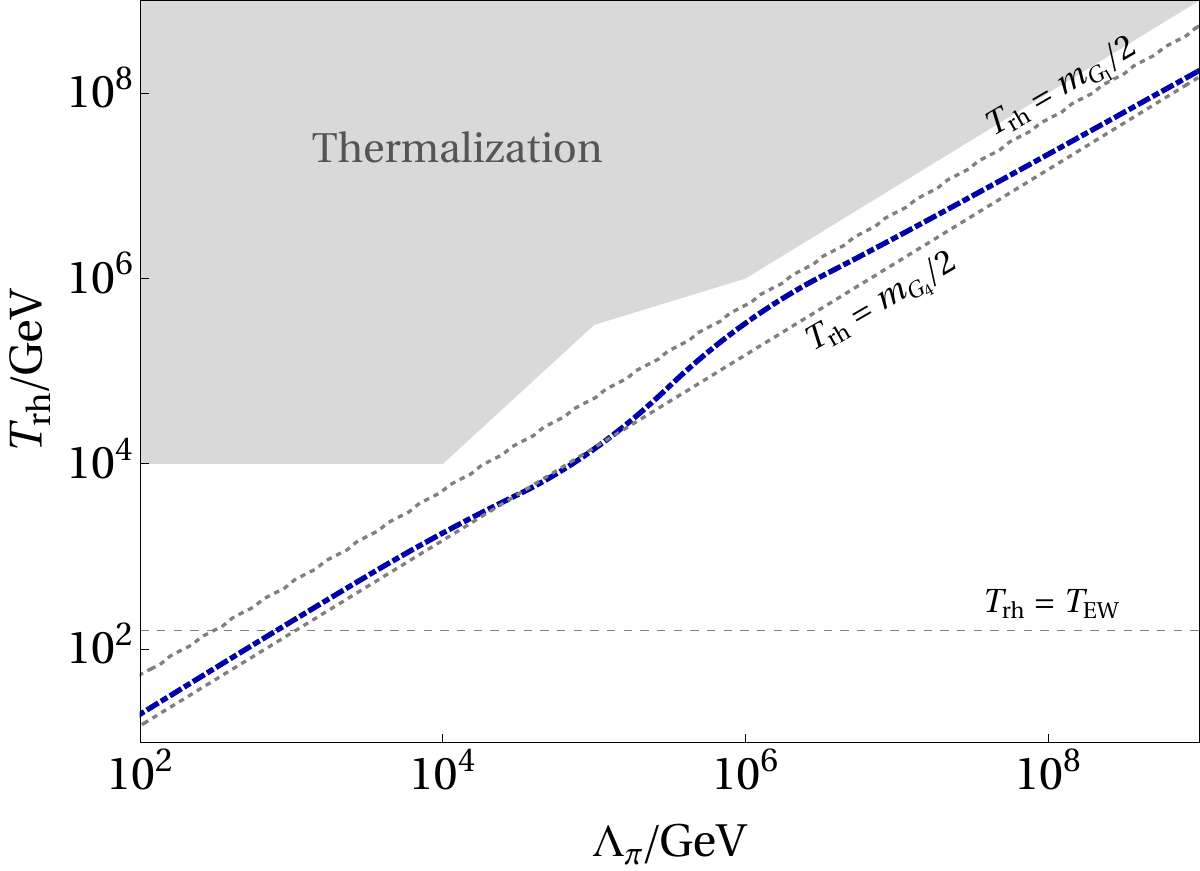}\\[10pt]
    \includegraphics[scale=0.45]{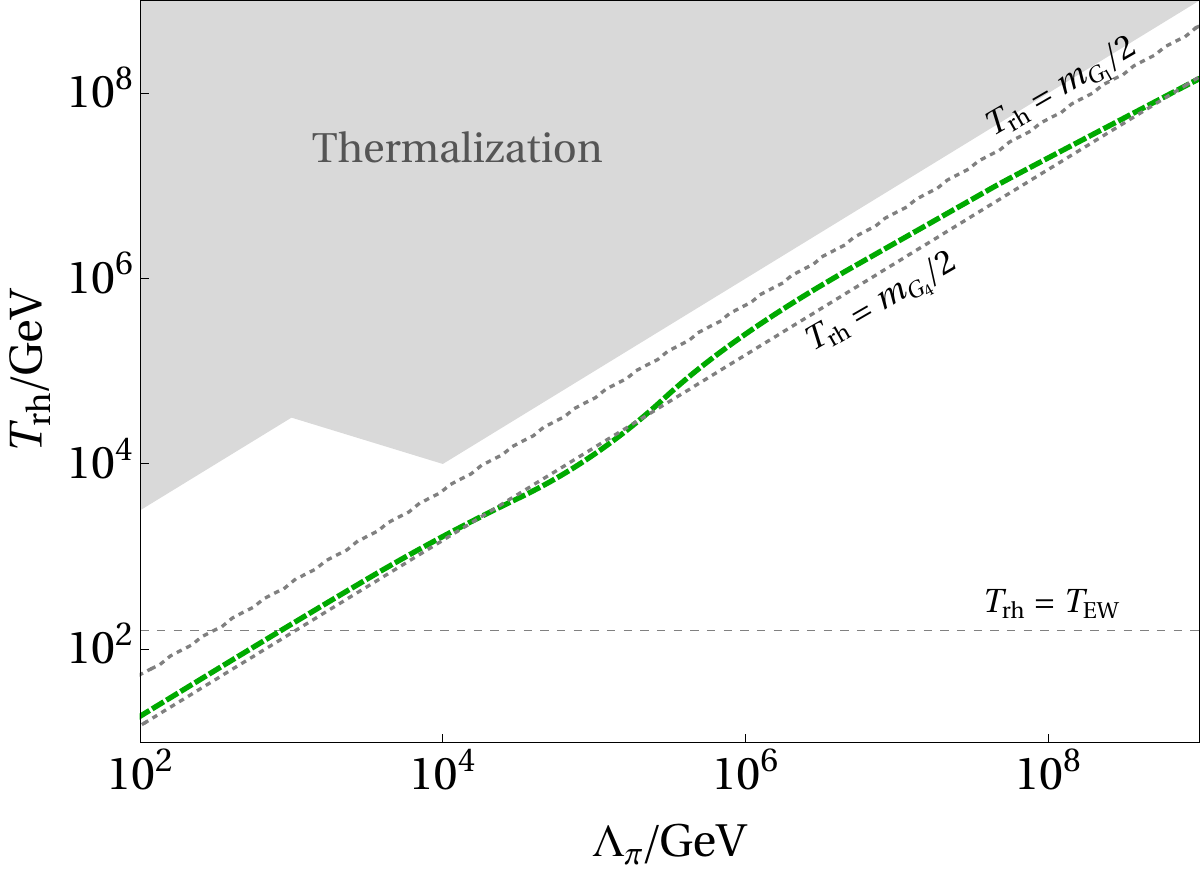}
    \caption{Relic density allowed parameter space in $\Trh-\Lambda_\pi$ plane, for spin-0 (top left), spin-1 (top right) and spin-1/2 (bottom) DM. The gray-shaded areas are the regions where chemical equilibrium with the bath is reached (and freeze-in does not occur). Here we have fixed $\mdm=10$ GeV and $C_{000}^{\phi\phi G}=10^{-12}\,\text{GeV}^{-1}$.}
    \label{fig:lamtrh}
\end{figure}

In Fig.~\ref{fig:relic} we illustrate the allowed parameter space for some specific choices of the DM mass, allowing different DM spins. Along the red, green and blue curves it is possible to produce all of the DM abundance for spin 0, 1/2 and spin-1 DM respectively, by varying the DM-graviton coupling $C_{000}^{\phi\phi G}\,\text{GeV}^{-1}$, taking all KK modes into account. As it is seen, larger coupling requires smaller $\Trh$ for a fixed DM mass, as the asymptotic DM yield has a typical UVFI characteristic, $Y_0\propto \mdm\,\Trh^{7(3)}\,\left(C_{000}^{\phi\phi G}\right)^2$, for $\Trh\ll(\gg) m_{G_n}/2$. The horizontal lines corresponding to $\Trh=m_{G_n}/2$ are added for comparison with approximate analytical results obtained in Eq.~\eqref{eq:dm-yield}. Irrespective of the DM spin, heavier DM requires lower $\Trh$ to produce the right abundance, for a given $C_{000}^{\phi\phi G}$. This is once again attributed to the UVFI feature, where bulk of the DM is produced at the maximum temperature of the Universe, which, in this case, is $\Trh$. Thus, to have the right abundance at a larger $\Trh$, it is required to have a lighter DM for a given coupling strength. Since $\Trh$ is the maximum bath temperature, it is therefore not possible to produce DM with mass exceeding $\Trh$. This constraint is more serious for heavier DM, as shown by the red shaded region in the bottom panel, where $\mdm>\Trh$. Finally, the gray shaded region is forbidden since it corresponds to DM thermalization, violating Eq.~\eqref{eq:dm-therm}. 

For a fixed DM of mass 10 GeV, we show the relic density allowed parameter space in Fig.~\ref{fig:lamtrh}. Irrespective of the DM spin, in all cases a higher reheating temperature is required as $\Lambda_\pi$ increases. This is, once again, because of the $\Trh/\Lambda_\pi^2$ dependence of the yield, following Eq.~\eqref{eq:yld-analyt}. As a result, a larger $\Trh$ results in overproduction of DM, that can be tamed down by choosing a higher $\Lambda_\pi$. The gray shaded regions are the regions where chemical  equilibrium with the bath is reached. Note that, in this case, since the DM mass $(\mdm=10\,\text{GeV})$ is well below the reheating temperature $(\Trh\gtrsim 100\,\text{GeV})$, hence the entire parameter space satisfies $\Trh>\mdm$.

Away from the sudden decay approximation for reheating, the bath temperature may rise to a temperature $\Tmax\gg\Trh$~\cite{Giudice:2000ex,Kolb:2003ke}. It is plausible that the DM relic density may be established during this reheating period, in which case its abundance will significantly differ from freeze-in calculations assuming radiation domination. In particular, it has been observed that if the DM is produced during the transition from matter to radiation domination via an interaction rate that scales like $\gamma(T)\propto T^n$, for $n>12$ the DM abundance is enhanced by a boost factor proportional to $(\Tmax/\Trh)^{n-12}$~\cite{Garcia:2017tuj}, whereas for $n\leq 12$ the difference between the standard UVFI calculation differ only by an $\mathcal{O}(1)$ factor from calculations taking into account non-instantaneous reheating. It has also been highlighted that the critical mass dimension of the operator at which the instantaneous decay approximation breaks down depend on the equation of state $\omega$, or equivalently, to the shape of the inflationary potential at the reheating epoch~\cite{Bernal:2019mhf,Garcia:2020eof,Co:2020xaf,Ahmed:2021fvt,Barman:2022tzk}. Therefore, the exponent of the boost factor becomes $(\Tmax/\Trh)^{n-n_c}$ with $n_c \equiv 6+2\, \left ( \frac{3-\omega}{1+\omega} \right ) $, showing a strong dependence on the equation of state~\cite{Bernal:2019mhf}. In the present case, as the interaction rate density $\gamma(T)\propto T^{12}$ for $T\ll m_{G_n}/2$ [cf. Eq.~\eqref{eq:gamma-analyt}], a sizeable boost factor is not expected, at least in the standard case where during reheating the inflaton energy density scales like non-relativistic matter. However, as the precise determination of such boost factors depends on the details of the reheating mechanism, in particular, the shape of the inflationary potential during reheating, it is beyond the scope of the present study. Given an inflationary model, a measurement (upper limit) on the tensor-to-scalar ratio $r$ can be translated into an upper bound on $\Trh$, thereby constraining the scale of inflation.

Precision measurements of primordial element abundances from Big Bang nucleosynthesis (BBN) suggest that the reheating temperature, $\Trh$, must be at least a few MeV~\cite{Sarkar:1995dd, Kawasaki:2000en,Hannestad:2004px, DeBernardis:2008zz, deSalas:2015glj,Hasegawa:2019jsa}. On the other hand, typical inflationary models predict an upper bound on $\Trh$ around \( 10^{16}~\mathrm{GeV} \) (see, e.g., Ref.~\cite{Linde:1990flp}). However, a high reheating temperature can potentially lead to issues with long-lived exotic relics that risk overclosing the Universe. A well-known example of this is the cosmological gravitino problem in supergravity scenarios~\cite{Moroi:1993mb}. As a result, supergravity models usually impose an upper limit on the reheating temperature of around \( 10^{10}~\mathrm{GeV} \), with even stricter constraints if the gravitino is light.
\subsection{Dark brane partially composite DM production}
In this scenario we consider a dark sector five-dimensional bulk field, which upon compactification can be Fourier expanded as,
\begin{align}
    \chi(x_\mu,\phi) = \sum_{n}\chi^{(n)}(x_\mu) f^{n}(\phi)\,.
\end{align}
The DM candidate is identified with KK-0 mode in the above expansion, that becomes dark brane partial composite or localized closer to the dark brane. Whereas, the KK partners of the DM, being $\sim \mathcal{O}(\text{TeV})$ heavy, are localized closer to the IR brane. This leads to an interesting situation where the KK partners of DM are in thermal equilibrium with the SM, as well as the KK partners of the SM. The interaction action for these KK modes with gravitons is similar to the action given in Eq.~\eqref{eq:bulkinteractiongraviton}. The five-dimensional action of the $m^{\rm th}$ and the $n^{\rm th}$ KK modes of DM with the KK modes of graviton ($h_{\mu\nu}^{(q)}$) is given by,
\begin{align}
& S= \sum_{m, n, q} \int d^5x \sqrt{g}
\,h_{\mu\nu}^{(q)}(x,y)\,T_{\rm DM}^{\mu\nu (m,n)}(x,y)   \nonumber\\&
=\sum_{m, n, q}\left\{\left[\int \frac{d y}{\sqrt{k}} \frac{e^{t kr_c |y|} \chi_{\rm DM}^{(m)} \chi_{\rm DM}^{(n)}\,\chi_G^{(q)}}{\sqrt{r_c}}\right] \frac{\kappa_4}{2}\right. \left.\times \int d^4 x\, \eta^{\mu\alpha}\,\eta^{\nu \beta}\,h_{\alpha\beta}^{(q)}(x)\,T_{\rm DM \ \mu \nu}^{(m, n)}\right\}\,,
\label{eq:bulkdminteractiongraviton}
\end{align} 
where $\chi_{\rm DM}^{(n)}$ and $T_{\rm DM \ \mu \nu}^{(m, n)}$ are the wave-profile in the bulk and the four-dimensional energy momentum tensor of the DM KK mode respectively. As shown here, these KK DM states can then decay to the DM zero mode via two/three body processes, as shown in Fig.~\ref{fig:feynKKdecay}, depending on the available phase space. The interaction term also allows for the DM$^{(n)}$ annihilation to SM particles through KK graviton mediation. Unlike the decay channel, the annihilation channel couplings are not suppressed making them crucial for computing DM$^{(n)}$ abundance.
\begin{figure}[htb!]
    \centering        
    \includegraphics[scale=0.12]{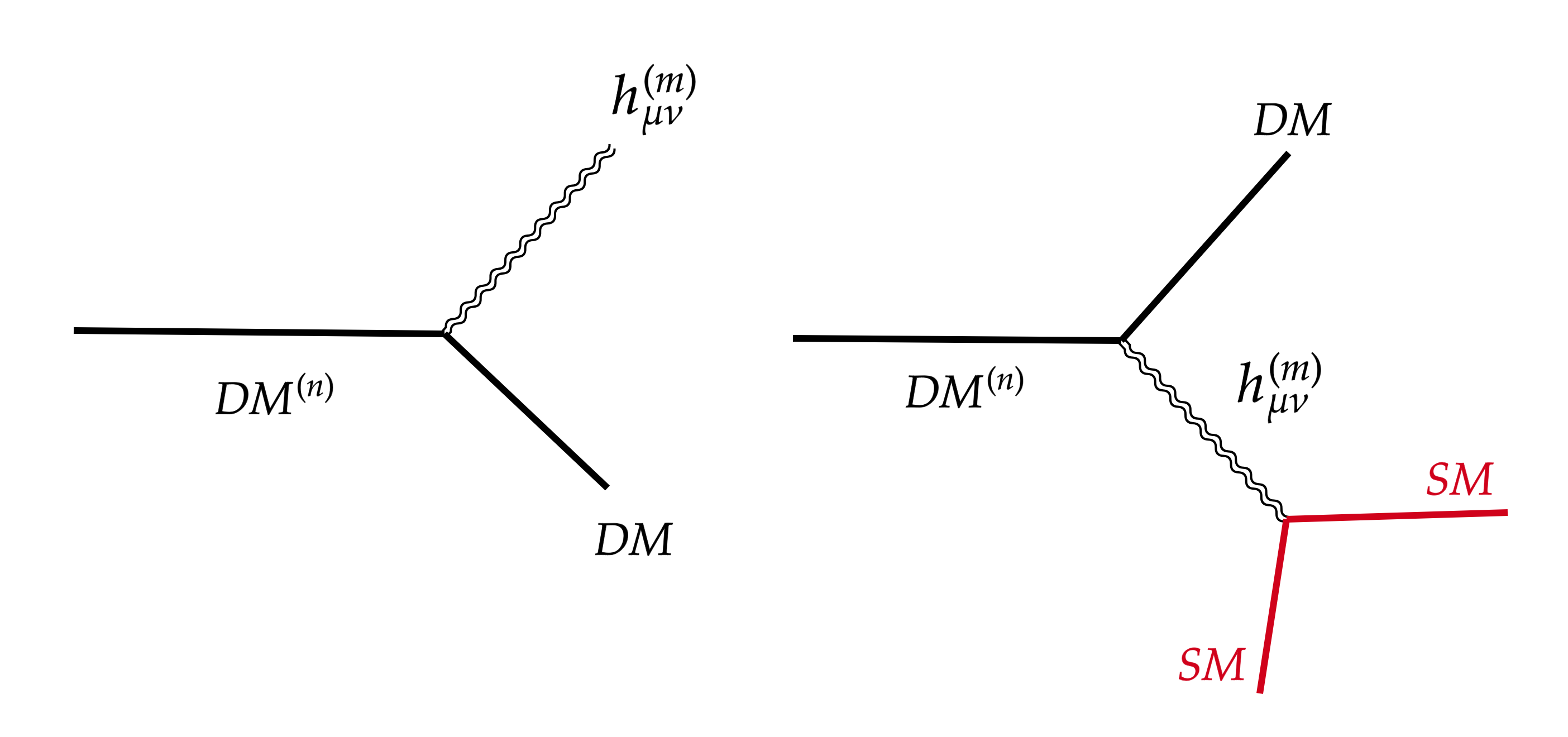}   
    \caption{Left: DM non-thermal production through the decay of $n^{\rm th}$ KK-DM partner via on-shell 2-body decay. Right: DM non-thermal production through off-shell decay of $n^{\rm th}$ DM KK-partner into the SM final states. }
    \label{fig:feynKKdecay}
\end{figure}

Due to the sizeable interaction rates with the thermal bath, DM$^{(n)}$ initially remain in thermal equilibrium in the early Universe. However, as their masses $\gtrsim 3$ TeV, they undergo early thermal freeze-out, typically resulting in a relatively large relic abundance. Subsequently, each DM$^{(n)}$ state can either annihilate themselves to SM particles via KK graviton mediation or decay into the lightest and stable DM state, denoted as DM$^{(0)}$. As a result, the final DM abundance is given by~\cite{Asaka:2006fs}
\begin{align}\label{eq:relDM0}
\Omega_{\rm DM^{(0)}}\,h^2 = \left(\frac{m_{\rm DM^{(0)}}}{m_{\rm DM^{(n)}}}\right)\,\Omega_{{\rm DM}^{(n)}}^{\rm FO}\,h^2\,,    
\end{align}
where $\Omega_{\rm DM^{(0)}}\,h^2$ is required to satisfy the observed DM abundance, and $m_{\rm DM^{(n)}}$ is the mass of the $n^\text{th}$ KK excitation of the DM. The problem arises because the freeze-out relic density $\Omega_{{\rm DM}^{(n)}}^{\rm FO}$ tends to be too large, potentially leading to an overclosed Universe, if the annihilation cross-section of DM $^{(n)}$+ DM $^{(n)}\to$ SM+ SM  is not efficient in reducing the DM$^{(n)}$ abundance. This resembles the ``superWIMP'' scenario~\cite{PhysRevLett.91.011302,Garny:2018ali}, commonly realized in models where the parent particle is part of a $Z_2$-odd dark sector. The $Z_2$ symmetry prevents its decay into SM particles, even though it may interact significantly with the SM. As a result, the parent particle undergoes freeze-out in a manner similar to a WIMP, while DM is subsequently produced from its decay—typically occurring much later in the evolution of the Universe.
\begin{figure}[htb!]
    \centering    \includegraphics[scale=0.5]{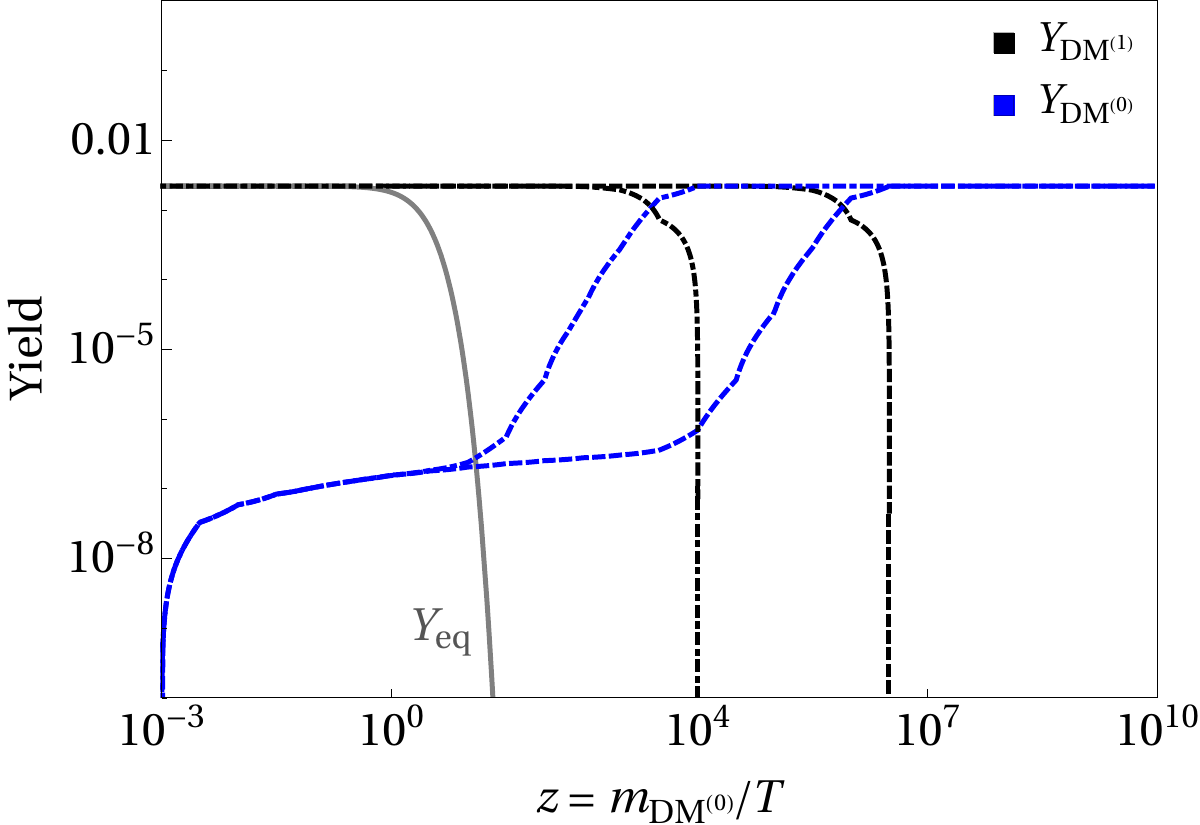}
    \caption{The evolution of DM$^{(1)}$ (in black) and DM$^{(0)}$ (in blue) yields as a function of the dimensionless quantity $z=m_{\rm DM^{(0)}}/T$. The gray solid curve corresponds to equilibrium yield. Here we have considered a scalar DM with mass $m_{\rm DM^{(0)}}=100$ GeV, $C_{000}^{\phi\phi G}=10^{-15}\,\text{GeV}^{-1}$ and $\langle\Gamma_{\rm DM^{(1)}\to DM^{(0)}}\rangle=\{10^{-25},\,10^{-20}\}$ GeV, shown via the dashed and dot-dashed curves, respectively.}
    \label{fig:coupled}
\end{figure}

To properly track the number densities of the DM components, we solve a set of coupled Boltzmann equations:
\begin{align}\label{eq:cbeq}
\frac{dY_{\rm DM^{(1)}}}{dz} &= -\frac{\mathfrak{s}(z)}{z\,H(z)}\Bigg[\langle\sigma v\rangle_{{\rm DM^{(1)}}\to \rm SM} \left(Y_{\rm DM^{(1)}}^2 - Y_{\rm DM^{(1)}}^{\rm eq\,2}\right) + \langle\sigma v\rangle_{{\rm DM^{(1)}} \to {\rm DM}^{(0)}}\,Y_{\rm DM^{(1)}}^2
\nonumber\\&
-\frac{z}{H(z)} \langle\Gamma_{\rm DM^{(1)} \to DM^{(0)}}\rangle\,Y_{\rm DM^{(1)}}\Bigg]\,, 
\nonumber\\
\frac{dY_{\rm DM^{(0)}}}{dz} &= \frac{\mathfrak{s}(z)}{z\,H(z)}\left[\frac{1}{\mathfrak{s}(z)} \langle\Gamma_{\rm DM^{(1)} \to DM^{(0)}}\rangle\,Y_{\rm DM^{(1)}} + \langle\sigma v\rangle_{{\rm DM^{(1)}} \to {\rm DM}^{(0)}}\,Y_{\rm DM^{(1)}}^2\right]\,,
\end{align}
where the equilibrium yield is given by
\[
Y^{\rm eq}(z) = \frac{45}{4\pi^4}\,\frac{g_{\rm DM}}{g_{*s}(z)}\,z^2\,K_2(z)\,; \quad z = \frac{m_{\rm DM^{(0)}}}{T}.
\]
The first line of Eq.~\eqref{eq:cbeq} takes care of DM$^{(1)}$ yield, where the first term in the square bracket is the annihilation of DM$^{(1)}$ into the SM final states, the second term is the annihilation into DM$^{(0)}$ final states, and the last term corresponds to the decay of DM$^{(1)}$. The second line of Eq.~\eqref{eq:cbeq} tracks the yield of DM$^{(0)}$. For illustration, we only consider the conversion DM$^{(1)} \to$ DM$^{(0)}$, and we assume a scalar DM candidate. Without delving into the detailed structure of the interactions, the thermally averaged decay rate is treated as a free parameter and is given by $\langle\Gamma_{\rm DM^{(1)} \to DM^{(0)}}\rangle = \left[K_1(z)/K_2(z)\right]\times\Gamma_{\rm DM^{(1)} \to DM^{(0)}}$. In this setup, the model is governed by three free parameters: $\{\Lambda_\pi,\, m_{\rm DM^{(0)}},\, C_{000}^{\phi\phi G}\}$, while the remaining quantities can be determined by fixing $ \Lambda_\pi$. 

The evolution of the DM yield is illustrated in Fig.~\ref{fig:coupled}, where the black dashed and dot-dashed curves show that DM$^{(1)}$ departs from equilibrium (gray solid curve) early due to Boltzmann suppression, leading to an overabundant stable DM population, as shown by the blue dashed and dot-dashed curves for two different benchmark choices of $\langle\Gamma_{\rm DM^{(1)} \to DM^{(0)}}\rangle$. An increased decay rate leads to an earlier freeze-in of DM$^{(0)}$, as all DM$^{(1)}$'s are decayed away earlier after they freeze-out. Note that, the final abundance of DM$^{(0)}$ follows the freeze-out abundance of DM$^{(1)}$, as one would expect from Eq.~\eqref{eq:relDM0}. This overabundance arises from the inefficiency of the annihilation channel DM$^{(1)}$ + DM$^{(1)} \to$ SM + SM, mediated by the KK gravitons. Though the couplings are not suppressed, the channel is far from resonance and hence does not provide enough cross-section for the DM$^{(n)}$'s to annihilate within a short period of time. However, this does not entirely rule out the possibility of a partially composite DM scenario within the present framework. If the mass spectrum were such that the annihilation process DM$^{(1)}$ + DM$^{(1)} \to$ SM + SM occurred near resonance, the resulting relic abundance could be significantly reduced. This opens up the possibility of reviving the scenario through a generalized model-building.
\section{Conclusions}
\label{sec:concl}
The persistent null results from all direct and indirect searches for dark matter (DM) have exposed plethora of issues and gaps in our understanding of a sector that comprises of $\sim 26\%$ of the universe. While the cosmological evidence for DM is robust, a concrete particle level description remains elusive. One intriguing framework is Planckian Interacting Dark Matter (PIDM), which posits DM candidates that interact solely via gravity, with interactions suppressed by the Planck scale. Such particles naturally evade detection in current experiments and may arise as composite states at or near the Planck scale. However, for PIDM to account for the observed CMB constraints, the reheating temperature needs to extremely large ($\gtrsim 10^{15}$ GeV). In this article, we consider a scenario in which the PIDM is a light composite particle with mass as low as 1 MeV. We demonstrate that the only consistent way to accommodate such a particle, while satisfying relic density constraints, is within a warped five-dimensional framework where it is possible to achieve lower reheating temperature, by tuning the the DM-graviton coupling. 

We adopt a five-dimension orbifolded $S^1/Z_2$ where, along with the {\it Ultra Violet} and {\it Infra Red} branes there also exist a {\it Dark} brane (DB) located in-between. To be consistent with PIDM framework, the DM is assumed to be localized to the {\it Dark} brane and it interacts only with the massless graviton and its heavier Kaluza Klein states with Planck suppressed coupling. To assure that, the DB is assumed to be located close to the UV brane. Heavy ($\gtrsim 100$ GeV) Standard Model (SM) fields and their KK modes, on the other hand, are localized close to the IR brane to satisfy the geometric Froggatt-Nielsen mechanism. Unlike the previously considered scenario of PIDM, here, an efficient annihilation channels of the heavy SM modes to DM emerges through the exchange of KK gravitons as shown in Fig.~\ref{fig:feyn}. While the geometry might look similar to few other DM constructions~\cite{Lee:2013bua,Lee:2024wes,Bernal:2020fvw} in warped five-dimensions, they assume the {\it Dark} brane to be close to IR ensuring strong interaction between DM and the heavy SM modes, thus straying away from the PIDM consideration. The present study also differs from~\cite{Lee:2013bua}, which considers DM production via freeze-out under the standard WIMP paradigm. Consequently, corresponding relic density allowed parameter space is tightly constrained from direct search experiments. In contrast,~\cite{Lee:2024wes} explores non-thermal freeze-in production, focusing primarily on the region where $\Trh < \mdm$, which they dub as the `low reheating' regime. However, such a scenario is not viable in our case due to the assumption of instantaneous reheating, that always demands $\Trh>\mdm$ in order for the bath to have enough energy to produce a pair of DM.

We consider two distinct realizations of DM in the warped PIDM setup and arrive at two main conclusions:
\begin{itemize}
\item First we take up the DB composite DM scenario, wherein the DM is is assumed to be localized on the dark brane. In this framework, the DM is sensitive to the maximal temperature attained by the Universe subsequent to reheating—a feature of ultraviolet (UV) freeze-in. Employing the observational constraint on the present-day relic abundance [cf. Fig.~\ref{fig:relic}], we derive bounds on the reheating temperature and the effective DM coupling. Our findings show that, it is possible to open up a large parameter space for the DM, over a mass range from MeV to TeV, with reheating temperature as low as 10 GeV to as large as $10^{10}$ GeV, depending on the strength of effective DM-graviton coupling that results in right DM abundance. For the purpose of this analysis, we adopt the assumption of instantaneous reheating, whereby the reheating temperature corresponds to the peak temperature achieved in the post-inflationary epoch. A complete analysis must take into account
the details of the inflationary model and the post-inflationary dynamics of the inflaton.

\item In the second scenario, the DM is assumed to be a partially composite state localized on the DB, with heavier KK excitations predominantly localized near the IR brane. These heavier KK modes are efficiently produced after inflation via their unsuppressed couplings to the massive KK gravitons, and they remain in thermal equilibrium with the radiation bath until the temperature of the Universe drops below their mass scale. Although these heavy KK states eventually decay into the lighter, stable DM candidate, their large initial abundance poses the risk of overclosing the Universe. In addition to decays, these KK modes can also undergo annihilation into SM particles via KK graviton exchange. If this annihilation proceeds near a resonance, it can sufficiently deplete their abundance, thereby ensuring consistency with the observed DM abundance.
\end{itemize}
\subsubsection*{Acknowledgments}
BB gratefully acknowledges the use of {\tt Package-X}~\cite{Patel:2016fam} for performing computations of decay and scattering amplitudes. M.T.A. acknowledges the financial support of DST through the INSPIRE Faculty grant DST/INSPIRE/04/2019/002507. The authors especially thank Rakesh Kumar S for the earlier collaboration and discussions.
\appendix
\section{Cross-sections \& decay rates}
\label{sec:cs}
\subsection{Graviton-mediated cross-sections}
\subsubsection{Spin-0 DM}
\label{sec:cs-0}
\begin{align}
& \sigma(s)_{H_nH_n\to \phi\phi}=\frac{s^3\,|S_{KK}|^2}{2880\,\pi\,\Lambda_\pi^2}\,\left(C_{000}^{\phi\phi G}\right)^2\,\left(1-x^2\right)^{5/2}\,\left(1-y^2\right)^{3/2}\,,
\nonumber\\&
\sigma(s)_{f_nf_n\to \phi\phi}=\frac{s^3\,|S_{KK}|^2}{180\,\pi}\,\left(C_{000}^{\phi\phi G}\,C_{mnq}^{ffG}\right)^2\,\left(1-x^2\right)^{5/2}\,\sqrt{1-y^2}\,\left(3+2\,y^2\right)\,,
\nonumber\\&
\sigma(s)_{V_nV_n\to \phi\phi}=\frac{s^3\,\big|S_{KK}\big|^2}{6480\,\pi}\,\left(C_{000}^{\phi\phi G}\,C_{mnq}^{VVG}\right)^2\,\left(1-x^2\right)^2\,\sqrt{\frac{1-x^2}{1-y^2}}\,\left(13+14\,y^2+3\,y^4\right)\,.
\end{align}
\subsubsection{Spin-1/2 DM}
\label{sec:cs-12}
\begin{align}
& \sigma(s)_{H_nH_n\to \psi\psi}=\frac{s^3\,|S_{KK}|^2}{180\,\pi\,\Lambda_\pi^2}\,\left(C_{000}^{\psi\psi G}\right)^2\,\left(1-x^2\right)^{3/2}\,\left(1-y^2\right)^{3/2}\,\left(3+2\,x^2\right)\,,
\nonumber\\&
\sigma(s)_{f_nf_n\to \psi\psi}=\frac{4s^3\,\left|S_{KK}\right|^2}{405\pi}\,\left(C_{000}^{\psi\psi G}\,C_{mnq}^{ffG}\right)^2\,(1-x^2)^{3/2}\,\sqrt{1-y^2}\,\left(1+\frac{2}{3}\,x^2\right)\,\left(1+\frac{2}{3}\,y^2\right)\,,
\nonumber\\&
\sigma(s)_{V_nV_n\to \psi\psi}=\frac{s^3\,|S_{KK}|^2}{405\,\pi}\,\left(C_{000}^{\psi\psi G}\,C_{mnq}^{VVG}\right)^2\,(1-x^2)\,\sqrt{\frac{1-x^2}{1-y^2}}\,\left(3+2x^2\right)\,\left(13+3y^4+14y^2\right)\,.
\end{align}
\subsubsection{Spin-1 DM}
\label{sec:cs-1}
\begin{align}
& \sigma(s)_{H_nH_n\to XX}=\frac{s^3\,|S_{KK}|^2}{2880\,\pi\,\Lambda_\pi^2}\,\left(C_{000}^{XXG}\right)^2\,\sqrt{1-x^2}\,\left(1-y^2\right)^{3/2}\,\left(13+3 x^4+14 x^2\right)\,,
\nonumber\\&
\sigma(s)_{f_nf_n\to XX}=\frac{s^3\,|S_{KK}|^2}{180\,\pi}\,\left(C_{000}^{XXG}\,C_{mnq}^{ffG}\right)^2\,\sqrt{\frac{1-x^2}{1-y^2}}\,\left(13+14\,x^2+3\,x^4\right)\,(1-y^2)\,\left(3+2\,y^2\right)\,,
\nonumber\\&
\sigma(s)_{V_nV_n\to XX}=\frac{s^3\,|S_{KK}|^2}{6480\,\pi}\,\left(C_{000}^{XXG}\,C_{mnq}^{VVG}\right)^2\,\,\sqrt{\frac{1-x^2}{1-y^2}}\,\left(13+3 x^4+14 x^2\right)\,\left(13+3 y^4+14 y^2\right)\,.
\end{align}
where $x=2\,\mdm/\sqrt{s}$ and $y_i=2\,m_i/\sqrt{s}$, $m_i$ being masses of different KK modes. Here, $S_{KK}$ corresponds to the KK-graviton propagators
\begin{align}
& S_{KK}=\sum_{n=0}^4\,\frac{1}{s-m_{G_n}^2+i\,\Gamma_{G_n}\,m_{G_n}}\,, 
\end{align}
where the propagator for a KK-graviton with mass $m_{G_n}$ reads
\begin{align}
& i\,\Delta_{{\mu\nu},{\rho\kappa}}(q)=\frac{iB_{\mu\nu,\rho\kappa}}{q^2-m_{G_n}^2+i\epsilon}\,,   
\end{align}
with
\begin{align}
& B_{\mu\nu,\rho\kappa} = -\frac{2}{3}\,\left(\eta_{\rho ,\kappa }-\frac{q_{\kappa } q_{\rho }}{m_{G_n}^2}\right) \left(\eta_{\mu ,\nu }-\frac{q_{\mu } q_{\nu }}{m_{G_n}^2}\right)+\left(\eta_{\nu ,\kappa }-\frac{q_{\kappa } q_{\nu }}{m_{G_n}^2}\right) \left(\eta_{\mu ,\rho }-\frac{q_{\mu } q_{\rho }}{m_{G_n}^2}\right)
\nonumber\\&
+\left(\eta_{\mu ,\kappa }-\frac{q_{\kappa } q_{\mu }}{m_{G_n}^2}\right) \left(\eta_{\nu ,\rho }-\frac{q_{\nu } q_{\rho }}{m_{G_n}^2}\right)\,.    
\end{align}
\subsection{Graviton Decay rates}
\subsubsection{To different KK-modes}
\begin{align}
& \Gamma_{H_n\,H_n}=\frac{m_{G_n}^3}{960\,\Lambda_\pi^2}\,\left(1-4\,r^2\right)^{5/2}~~\text{(spin-0)}
\nonumber\\&
\Gamma_{f_n\,f_n}=\frac{\left(C_{mnq}^{ffG}\right)^2\,m_{G_n}^3}{15\,\pi}\,\left(1-4r^2\right)^{3/2}\,\left(3+8r^2\right)~~\text{(spin-1/2)}
\nonumber\\&
\Gamma_{V_n\,V_n}=\frac{\left(C_{mnq}^{VVG}\right)^2\,m_{G_n}^3}{240\,\pi}\,\left(1-4\,r^2\right)^{1/2}\,\left(13+56\,r^2+48\,r^4\right)~~\text{(spin-1)}\,,
\end{align}
where $r=m_i/m_{G_n}$. 
\subsubsection{To different DM spins}
\begin{align}
& \Gamma_{\phi\,\phi}=\frac{\left(C_{000}^{\phi\phi G}\right)^2\,m_{G_n}^3}{960}\,\left(1-4\,r^2\right)^{5/2}~~\text{(spin-0)}
\nonumber\\&
\Gamma_{\psi\,\psi}=\frac{\left(C_{000}^{\psi\psi G}\right)^2\,m_{G_n}^3}{15\,\pi}\,\left(1-4r^2\right)^{3/2}\,\left(3+8r^2\right)~~\text{(spin-1/2)}
\nonumber\\&
\Gamma_{XX}=\frac{\left(C_{000}^{XXG}\right)^2\,m_{G_n}^3}{240\,\pi}\,\left(1-4\,r^2\right)^{1/2}\,\left(13+56\,r^2+48\,r^4\right)~~\text{(spin-1)}\,,
\end{align}
where $r=\mdm/m_{G_n}$.
\bibliography{ref}
\bibliographystyle{JHEP}
\end{document}